\documentclass[prd,nofootinbib,preprint,superscriptaddress]{revtex4-2}
\pdfoutput=1

\usepackage{graphicx}
\usepackage{epsfig}
\usepackage{amsmath}
\usepackage{amsfonts}
\usepackage{amssymb}
\usepackage{color}
\usepackage[bottom]{footmisc}
\usepackage{array,multirow,makecell}
\usepackage{slashed}
\usepackage[colorlinks,citecolor=blue]{hyperref}
\usepackage{float}
\usepackage{ulem}
\usepackage[utf8]{inputenc}
\usepackage[english]{babel}
\usepackage{url}
\usepackage{hyperref}
\usepackage{xcolor}
\usepackage{subfigure}
\usepackage{lipsum}
\usepackage[bottom]{footmisc}
\usepackage{pifont}
\newcommand{\baz}{\begin{array}{cc}}

\newcommand{\lhp}{\lambda_{H\varphi}}

\newcommand{\mdm}{m_\chi}
\newcommand{\ydm}{y_\chi}
\newcommand{\pmns}{U_\text{PMNS}}
\begin{document}
\title{Probing Freeze-in Dark Matter via Heavy Neutrino Portal}
\author{Basabendu Barman}
\email{basabendu88barman@gmail.com}
\affiliation{Centro de Investigaciones, Universidad Antonio Nariño\\
Carrera 3 este \# 47A-15, Bogotá, Colombia}
\affiliation{Institute of Theoretical Physics, Faculty of Physics, University of Warsaw,\\ ul. Pasteura 5, 02-093 Warsaw, Poland}

\author{P. S. Bhupal Dev}
\email{bdev@wustl.edu}
\affiliation{Department of Physics and McDonnell Center for the Space Sciences, \\ Washington University, St. Louis, MO 63130, USA}
\affiliation{Theoretical Physics Department, Fermilab, P.O. Box 500, Batavia, IL 60510, USA}

\author{Anish Ghoshal}
\email{anish.ghoshal@fuw.edu.pl}
\affiliation{Institute of Theoretical Physics, Faculty of Physics, University of Warsaw,\\ ul. Pasteura 5, 02-093 Warsaw, Poland}

\begin{abstract}
We explore the possibility of probing freeze-in dark matter (DM) produced via the right-handed neutrino (RHN) portal using the RHN search experiments. We focus on a simplified framework of minimally-extended type-I seesaw model consisting of only four free parameters, namely the RHN mass, the fermionic DM mass, the Yukawa coupling between the DM and the RHN, and a real singlet scalar mass. We consider two cases for the DM production either via decay of the thermal RHN or via scattering of the bath particles mediated by the RHN. In both cases, we show that for sub-TeV scale DM masses, the allowed model parameter space satisfying the observed DM relic density for freeze-in scenario falls within the reach of current and future collider, beam dump and forward physics facilities looking for feebly-coupled heavy neutrinos. 
\end{abstract}
\begin{flushright}
  PI/UAN-2022-718FT \\
\end{flushright}
\maketitle
\section{Introduction}
\label{sec:intro}
The nature of dark matter (DM) remains mysterious and one of the most important questions in fundamental physics today~\cite{ParticleDataGroup:2020ssz}. On one hand, we have overwhelming evidence of DM as the dominant matter component of our Universe today from a plethora of cosmological and astrophysical observations~\cite{Bertone:2004pz}. On the other hand, there is no firm evidence for DM coupling to the Standard Model (SM) sector except via gravity. No matter how non-gravitational DM interactions may manifest, it would require some beyond the SM (BSM) physics to provide a suitable particle DM candidate~\cite{Feng:2010gw}.    

Among various possible DM candidates, the Weakly Interacting Massive Particle (WIMP)~\cite{Steigman:1984ac} paradigm has gained a lot of attention so far, thanks to its miraculous property of being able to reproduce the observed relic abundance via weak-scale interaction cross-sections for a wide range of DM masses~\cite{Jungman:1995df, Steigman:2012nb}. In spite of being so appealing, the strong experimental constraints (from direct detection, indirect detection and colliders) on the typical WIMP parameter space~\cite{Roszkowski:2017nbc, Arcadi:2017kky} have recently motivated quests for DM beyond the standard WIMP paradigm~\cite{Cooley:2022ufh}. 

Since DM is electrically neutral, a simple alternative to the WIMP paradigm (where the DM is typically the neutral component of an electroweak multiplet; see e.g.~Ref.~\cite{Bottaro:2021snn}) is to have the DM as a pure singlet under the SM gauge group. In this case, the DM can interact with the SM sector only via the so-called `portals'. There exist only three such portals in the SM, depending on whether the mediator has spin-0 (Higgs portal)~\cite{Silveira:1985rk, McDonald:1993ex, Burgess:2000yq, Patt:2006fw, Fedderke:2014wda, Gross:2015cwa, DiBari:2016guw, Casas:2017jjg, Bauer:2017fsw, DeRomeri:2020wng, Arcadi:2021mag, Lebedev:2021xey,Ghosh:2023tyz}, spin-1 (vector portal)~\cite{Galison:1983pa, Holdom:1985ag, Pospelov:2007mp, Arkani-Hamed:2008hhe, Mambrini:2010dq, Dudas:2013sia,Alves:2013tqa, Lebedev:2014bba,Alves:2015mua, Okada:2016tci, Fitzpatrick:2020vba, CarrilloGonzalez:2021lxm}, or spin-1/2 (neutrino portal)~\cite{Pospelov:2007mp,Falkowski:2009yz, Cherry:2014xra,GonzalezMacias:2015rxl,Gonzalez-Macias:2016vxy, Escudero:2016tzx,Escudero:2016ksa,Tang:2016sib, Batell:2017rol, Batell:2017cmf, Folgado:2018qlv, Becker:2018rve, Chianese:2018dsz, Berlin:2018ztp,  Bandyopadhyay:2018qcv,  Bian:2018mkl, Chianese:2019epo, Chianese:2020khl, Lamprea:2019qet, Blennow:2019fhy,Patel:2019zky,Cosme:2020mck,Bandyopadhyay:2020qpn, Biswas:2021kio, Cheng:2021umr,Borah:2021pet,Paul:2021ewd,Hufnagel:2021pso, Biswas:2022vkq,Liu:2022rst}. In this paper, we will focus on the neutrino portal scenario which is particularly interesting because of its intimate connection to neutrino mass -- another outstanding puzzle that also calls for some BSM physics~\cite{Mohapatra:2006gs}.    

A simple realization of the neutrino portal relies on DM interactions being mediated by SM gauge-singlet right-handed neutrinos (RHNs), also known as the sterile neutrinos or heavy neutral leptons in the literature. The RHNs are well motivated from the type-I seesaw mechanism for neutrino mass generation~\cite{Minkowski:1977sc, Mohapatra:1979ia, Gell-Mann:1979vob, Yanagida:1979as, Glashow:1979nm, Schechter:1980gr}. Depending on their mass and Dirac Yukawa couplings, which together determine their mixing with the SM neutrinos, the RHNs can be searched for in a wide range of experiments, such as beta decay, meson decay, beam dump, and colliders; for a comprehensive summary of the existing constraints and future prospects of RHN searches, see e.g.~Refs.~\cite{Atre:2009rg, Bolton:2019pcu, Abdullahi:2022jlv}. In this paper, we show that the same RHN parameter space that can be probed in future experiments can also reproduce the observed DM relic 
density, if the RHNs are the only mediators between the SM and the DM sectors. In addition, we will assume that the portal couplings to the dark sector are sufficiently small so that the DM never reaches chemical equilibrium with the thermal bath. In this case, the DM is slowly populated in the Universe by either decay or annihilation processes involving the RHNs, until the production ceases due to Boltzmann suppression as the Hubble temperature drops below the RHN mass. Therefore, this is a freeze-in, or feebly interacting massive particles (FIMP) DM scenario~\cite{Hall:2009bx, Bernal:2017kxu}, in contrast with the freeze-out scenario for WIMPs. Due to their tiny interaction strength with the visible sector, FIMPs are inherently very difficult to search for directly in conventional DM direct detection, indirect detection, or collider experiments\footnote{Direct detection prospects of FIMP-like DM have been discussed in, for example, Refs.~\cite{Hambye:2018dpi,Elor:2021swj}.}. However, unlike freeze-out, for freeze-in one typically looks for signatures of the portal itself and its associated tiny couplings. For instance, the feeble couplings associated with the portal could make either the heavier dark sector particles or the mediator itself long-lived, leading to signatures in lifetime and intensity frontier experiments~\cite{Belanger:2018sti}. Other examples involving properties of the individual BSM models like kinetic mixing~\cite{Hambye:2018dpi}, temperature corrections~\cite{Darme:2019wpd,Biondini:2020ric} and scale-invariance~\cite{Barman:2021lot,Barman:2022njh} have been proposed for freeze-in mechanism that can be searched for in direct detection experiments as well. Similarly, a non-standard cosmological era can also make freeze-in sensitive to indirect detection~\cite{Cosme:2020mck}. 

In this paper, we show that the RHN portal effectively provides a complementary laboratory probe of the FIMP DM scenario. Although we study the minimal type-I seesaw for concreteness, our prescription for RHN-portal searches is generic and can also be applied to other neutrino mass models, such as inverse seesaw~\cite{Mohapatra:1986bd} and radiative models~\cite{Cai:2017jrq, Babu:2019mfe}, as well as to other dark singlet fermion portal models (see e.g.~Refs.~\cite{Darme:2020ral, Chianese:2021toe, Coy:2021sse}, and references therein). The main novelty of our analysis is the projection of FIMP DM-allowed parameter space onto the RHN mass-mixing plane, which makes it straightforward to correlate the RHN-portal freeze-in parameter space with the experimental detection prospects at the RHN-frontier.  

The rest of the paper is organized as follows. In Sec.~\ref{sec:model} we have introduced the details of the RHN portal freeze-in DM model under consideration. We then discuss the DM phenomenology in Sec.~\ref{sec:fi-dm}, where we elucidate the sensitivity reach of present and future experiments in probing the relic density allowed parameter space, possible collider search prospects for this model is discussed in Sec.~\ref{sec:collider}, and finally we conclude in Sec.~\ref{sec:concl}. Freeze-in reaction densities are presented in  Appendix~\ref{sec:app-reacden}, relevant RHN decay widths, together with DM production cross-sections are listed in Appendix~\ref{sec:app-RHN-decay} and Appendix~\ref{sec:app-22-dm} respectively, finally, production cross-sections for $\varphi$ are reported in Appendix~\ref{app:phi-prod}.
\section{The model}
\label{sec:model}
We extend the SM particle content with the addition of the following:
\begin{itemize}
    \item SM gauge-singlet RHNs $N_i$. We need at least two RHNs (i.e., $i=1,2$) in order to reproduce two nonzero mass-squared differences, as observed in neutrino oscillation data, using the seesaw mechanism. For our current interest, a hierarchical spectrum can be assumed, so that only the lightest RHN $N_1$ will be relevant for us.\footnote{If the RHN mass is the keV range and its Yukawa couplings are sufficiently small, then it could be a DM candidate itself~\cite{Asaka:2005an,Drewes:2016upu,Datta:2021elq}, but here we are interested in the Yukawa couplings relevant for seesaw and potentially accessible in laboratory experiments.}  
    \item A gauge-singlet Majorana fermion $\chi$ which serves as the DM candidate. Note that a Dirac fermion would also serve the purpose, but at the expense of doubling the degrees of freedom. 
    \item A real singlet scalar $\varphi$ which is needed to connect the DM to the RHN portal.   
\end{itemize}
We will assume that both $\chi$ and $\varphi$ are charged under a $Z_2$ symmetry and that $\chi$ is lighter than $\varphi$ to ensure the stability of the DM. The SM particles and the RHNs are assumed to be even under this $Z_2$ symmetry, which forbids couplings between SM and dark sector particles ($\chi$, $\varphi$). 
The relevant piece of the Lagrangian giving rise to neutrino mass is given by 
\begin{equation}
-\mathcal{L}_\nu = (Y_D)_{\alpha j}\,\overline{L}_\alpha\, H \,N_j +\frac{1}{2}(M_N)_{ij}\,\overline{N^c_i}\,N_j+\text{H.c.} \,,
\label{eq:lnu}
\end{equation}
where $L$ and $H$ are the $SU(2)_L$ lepton and Higgs doublets respectively, and $\alpha=e,\mu,\tau$ is the flavor index. The interaction Lagrangian for the dark sector containing the singlet Majorana DM $\chi$ and the real singlet scalar $\varphi$ reads
\begin{equation}
-\mathcal{L}_\text{dark}= \ydm\,\overline{N^c}\,\varphi\,\chi+m_\chi\,\overline{\chi^c}\,\chi
 +V(H,\varphi)
+\text{H.c.}\,,
\label{eq:ldm}
\end{equation}
where $V(H,\varphi)$ is the scalar potential (see below) and we have assumed a universal coupling of DM to the RHNs. The RHNs serve as the portal to mediate the interactions between the dark and visible sectors, owing to the couplings $Y_D$ and $y_\chi$. Note that the same $Y_D$ is also involved in active-sterile neutrino mixing, leading to light neutrino mass generation via type-I seesaw mechanism. 

Once the SM Higgs doublet gets a nonzero vacuum expectation value (VEV)
\begin{equation}
H = \frac{1}{\sqrt{2}}\,\begin{pmatrix}
0 \\ h+v
\end{pmatrix}\,    
\end{equation}
with $v\simeq 246$ GeV, we obtain the Dirac mass matrix $M_D=Y_D\,\langle H\rangle$. The singlet scalar $\varphi$, on the other hand, does not acquire a VEV, and therefore, there is no mixing between the DM and the RHNs. The Lagrangian in Eq.~\eqref{eq:lnu} in the flavor basis then reads
\begin{equation}
-\mathcal{L}_\nu = \frac{1}{2}\,
\begin{pmatrix}
\overline{(\nu_L)^c} & \overline{N}
\end{pmatrix}\,\mathcal{M}\,
\begin{pmatrix}
\nu_L\\ \ N^c
\end{pmatrix}+\text{H.c.}\,,
\end{equation} 
where the mass matrix can be realized as
\begin{equation}
\mathcal{M}=\begin{pmatrix}
0 & M_D \\ M_D^T & M_N
\end{pmatrix}\,,    
\end{equation}
that can be diagonalized using a unitary matrix $\mathcal{U}$: 
\begin{equation}
\mathcal{M}_\text{diag} = \mathcal{U}^T.\,\mathcal{M}.\,\mathcal{U}\,,    
\end{equation}
obtaining masses of the neutrinos in the physical basis. We work in a basis where $M_N$ is diagonal, i.e., $\widehat{M}_N\equiv M_N={\rm diag}(M_1,M_2)$, and express the Yukawa matrix following the Casas-Ibarra (CI)  parametrization~\cite{Casas:2001sr} as
\begin{equation}\label{eq:CI}
Y_D = \frac{\sqrt{2}}{v}\,\sqrt{\widehat{M}_N}\,\mathbb{R}\,\sqrt{\widehat{m}_\nu}\,\pmns^\dagger\,,    
\end{equation}
with $\pmns$ being the PMNS matrix that diagonalizes the active neutrino sector (ignoring any non-unitarity effects) $\widehat{m}_\nu={\rm diag}(m_1,m_2,m_3)$, while $\mathbb{R}$ is an arbitrary complex orthogonal rotation matrix with $\mathbb{R}^T\,\mathbb{R}=\mathbb{I}$. In the minimal seesaw scenario with two RHNs~\cite{Frampton:2002qc}, considering either normal hierarchy (NH) or inverted hierarchy (IH) amongst the light neutrino masses, one can define the rotation matrices accordingly as
\begin{align}
&\mathbb{R}_\text{NH} =
\begin{pmatrix}
0 & \cos z & \sin z \\
0 & -\sin z & \cos z 
\end{pmatrix},\, \qquad 
\mathbb{R}_\text{IH} =
\begin{pmatrix}
\cos z & -\sin z & 0 \\
\sin z & \cos z & 0 
\end{pmatrix}\, ,
\end{align}
where $z$ is in general a complex angle. This choice automatically implies that the lightest active neutrino is massless. The mass eigenstates can be defined via the unitary rotation: 
\begin{equation}
\begin{pmatrix}
\nu_L \\ N^c
\end{pmatrix}=\mathcal{U}\,
\begin{pmatrix}
\nu_{i} \\ N_{j}
\end{pmatrix}\,,
\end{equation}
and the matrix $\mathcal{U}$ can be expressed as (expanding in terms of $M_DM_N^{-1}$)
\begin{equation}
\mathcal{U}=
\begin{pmatrix}
U_{\nu\nu} & U_{\nu N} \\ U_{N\nu} & U_{NN}
\end{pmatrix} 
\, .
\end{equation}
In terms of the CI parametrization [cf.~Eq.~\eqref{eq:CI}], to leading order, we find~\cite{He:2009ua,Cosme:2020mck} 
\begin{align}\label{eq:umatrix}
& U_{\nu\nu} \approx \pmns\nonumber
\nonumber\\&
U_{\nu N}\approx M_D^\dagger\,M_N^{-1}= \sqrt{2}\,\pmns\,\sqrt{\hat{m}_\nu}\,\mathbb{R}^\dagger\,M_N^{-1/2}\, , 
\nonumber\\&
U_{N\nu}\approx -M_N^{-1}\,M_D\,U_{\nu\nu} = -\sqrt{2}\,M_N^{-1/2}\,\mathbb{R}\,\sqrt{\hat{m}_\nu}\, .
\nonumber\\&
U_{NN}\approx\mathbb{I}\,,
\end{align}
The charged current interaction vertices are then modified as  
\begin{equation}
\mathcal{L}_{\rm CC} \supset\frac{g}{\sqrt{2}}\,\Bigl[(U_{\nu\nu})_{\alpha i}\, \overline{\ell}_{L\alpha}\,\gamma^\mu\,\nu_{i}+(U_{\nu N})_{\alpha j}\,\overline{\ell}_{L\alpha}\,\gamma^\mu\,N_{j}\,+\text{H.c.}\Bigr]\,W_\mu^-\,
\end{equation}
($g$ being the $SU(2)_L$ gauge coupling strength), whereas the neutral current interaction vertices are modified as 
\begin{align}
&\mathcal{L}_{\rm NC} \supset\frac{g}{2\,\cos\theta_w}\,\Bigl[(U_{\nu\nu}^\dagger\,U_{\nu\nu})_{ij}\,\overline{\nu}_{i}\,\gamma^\mu\,\nu_{j}+(U_{\nu N}^\dagger\,U_{\nu N})_{kl}\,\overline{N}_{k}\,\gamma^\mu\,N_{l}\nonumber\\&+(U_{\nu N}^\dagger\,U_{\nu\nu})_{ki}\,\overline{N}_{k}\,\gamma^\mu\,\nu_{i}+(U_{\nu\nu}^\dagger\,U_{\nu N})_{ik}\,\overline{\nu}_{i}\,\gamma^\mu\,N_{k}\Bigr]\,Z_\mu\,, 
\end{align}
where $\theta_w$ is the weak mixing angle and we have utilized $U_{\nu N}^T\,M_D=\widehat{m}_\nu\,U_{\nu\nu}^\dagger$ and $U_{NN}^T\,M_D=M_N\,U_{\nu N}^\dagger$. Similarly, the DM-neutrino interaction Lagrangian can be written in the physical basis as
\begin{align} &-\mathcal{L}_\text{dark}'\supset 
 \ydm\,\sum_k\left[(U_{N\nu}^T)_{ki}\,\overline{\nu}_i\,\varphi\,\chi+
(U_{NN}^T)_{kj}\,\overline{N}_j\,\varphi\,\chi\right]+\text{H.c}\,.
\end{align}

The renormalizable scalar potential [cf.~Eq.~\eqref{eq:ldm}] involving the two scalars of the theory, namely, $\{\varphi,H\}$ is given by
\begin{equation}
V(H,\varphi)=-\mu_H^2\,(H^\dag H)+\lambda_H\,(H^\dag H)^2+ \mu_\varphi^2\,\varphi^2+\lambda_\varphi\,\varphi^4+\lambda_{H\varphi}\,\varphi^2\,(H^\dag H)\,.
\label{eq:pot}
\end{equation}
After electroweak symmetry breaking, the scalar mass matrix is given by
\begin{equation}
\mathbb{M}^2=\left(
\begin{array}{cc}
 2\,v^2\,\lambda_H & 0 \\
 0 & \lambda_{H\varphi}\,v^2+2\,\mu_\varphi^2 \\
\end{array}
\right)\,,    
\end{equation}
where the two terms can be identified as the squared mass of the SM Higgs $h$ and the singlet $\varphi$ respectively. 

\section{Freeze-in production of Dark Matter}
\label{sec:fi-dm}
In the present set-up the DM can be produced via: (a) on-shell decay of RHN if $M_N>\mdm+m_\varphi$, or (b) 2-to-2 scatterings mediated by RHN. In case where the decay is present, the scattering will be sub-dominant (which we shall show in a moment), hence in the following we treat DM production from decay and that from scattering separately. We assume that the lightest RHN $N_1$ is responsible for freeze-in production of DM. Now, since freeze-in happens when thermal bath species produce DM via out-of-equilibrium processes, it is important to know whether the RHNs thermalize with the SM particles during the freeze-in production. As argued in Ref.~\cite{Hernandez:2013lza,Hernandez:2014fha}, in low-scale type-I seesaw models with three extra sterile states, full thermalization in the early Universe is always reached for three RHN states if the lightest neutrino mass is above $\mathcal{O}(10^{-3})$ eV, while if the lightest neutrino mass is below $\mathcal{O}(10^{-3})$ eV, only one of the sterile states might never thermalize. Thus in the subsequent analysis we will consider the freeze-in production to be happening from the thermal bath containing the RHNs. The principal assumption for freeze-in is to consider that DM abundance was zero in the early Universe and then the DM is produced gradually from the thermal bath with time via feeble renormalizable interactions -- the so-called infrared (IR) freeze-in mechanism; see e.g.~Refs.~\cite{McDonald:2001vt,Hall:2009bx,Chu:2011be,Blennow:2013jba, Dev:2013yza, Biswas:2018aib, Heeba:2018wtf, Biswas:2019iqm, Barman:2021lot}.\footnote{This is in contrast with the ultraviolet freeze-in, where the dark and visible sectors are coupled only via non-renormalizable operators~\cite{Elahi:2014fsa}.} 

\begin{figure}[htb!]
    \centering
    \includegraphics[scale=0.35]{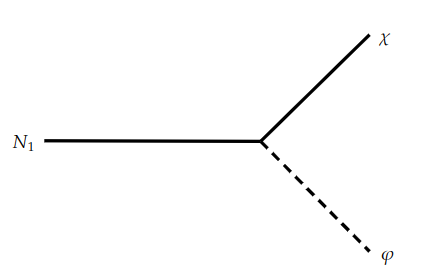}\\[10pt]
    \includegraphics[scale=0.5]{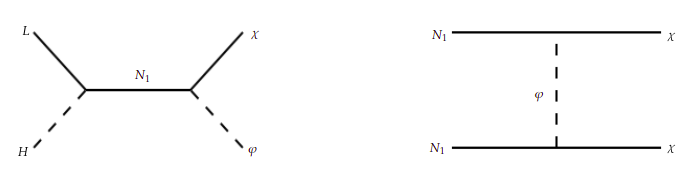}
    \caption{Relevant dark matter production channels via 1-to-2 decay (top) 2-to-2 scattering processes (bottom) involving the RHNs.}
    \label{fig:DM-prod}
\end{figure}

In the present scenario following are the relevant DM production channels:
\begin{itemize}
\item Decay: $N_1\to\chi\varphi$ 
\item $s$-channel RHN mediated scattering: $L\,H\to\chi\varphi\,,  V\,L\to\chi\varphi$ with $V\in W^\pm\,,Z$.
\item $t$-channel $\varphi$ mediated scattering: $N_1\,N_1\to\chi\,\chi$.
\end{itemize}
The corresponding Feynman graphs are shown in Fig.~\ref{fig:DM-prod}. 

Now, the Boltzmann equation (BE) governing the DM number density can be written in terms of the DM yield defined as a ratio of the DM number density to the entropy density in the visible sector, i.e.,  $Y_\chi=n_\chi/s$. The BE can then be expressed in terms of the reaction densities as
\begin{equation}\label{eq:beq}
x\,H\,s\,\frac{dY_\chi}{dx} = \gamma_\text{ann}+\gamma_\text{decay}\,,
\end{equation}
where $x\equiv\mdm/T$ is a dimensionless quantity. The complete expressions for the reaction densities $\gamma_i$'s can be found in Appendix~\ref{sec:app-reacden}. Since we investigate a feebly coupled sector, the back reactions in the DM production processes can be neglected~\cite{Hall:2009bx}.

\begin{figure}[htb!]
    \centering
    \includegraphics[width=0.495\textwidth]{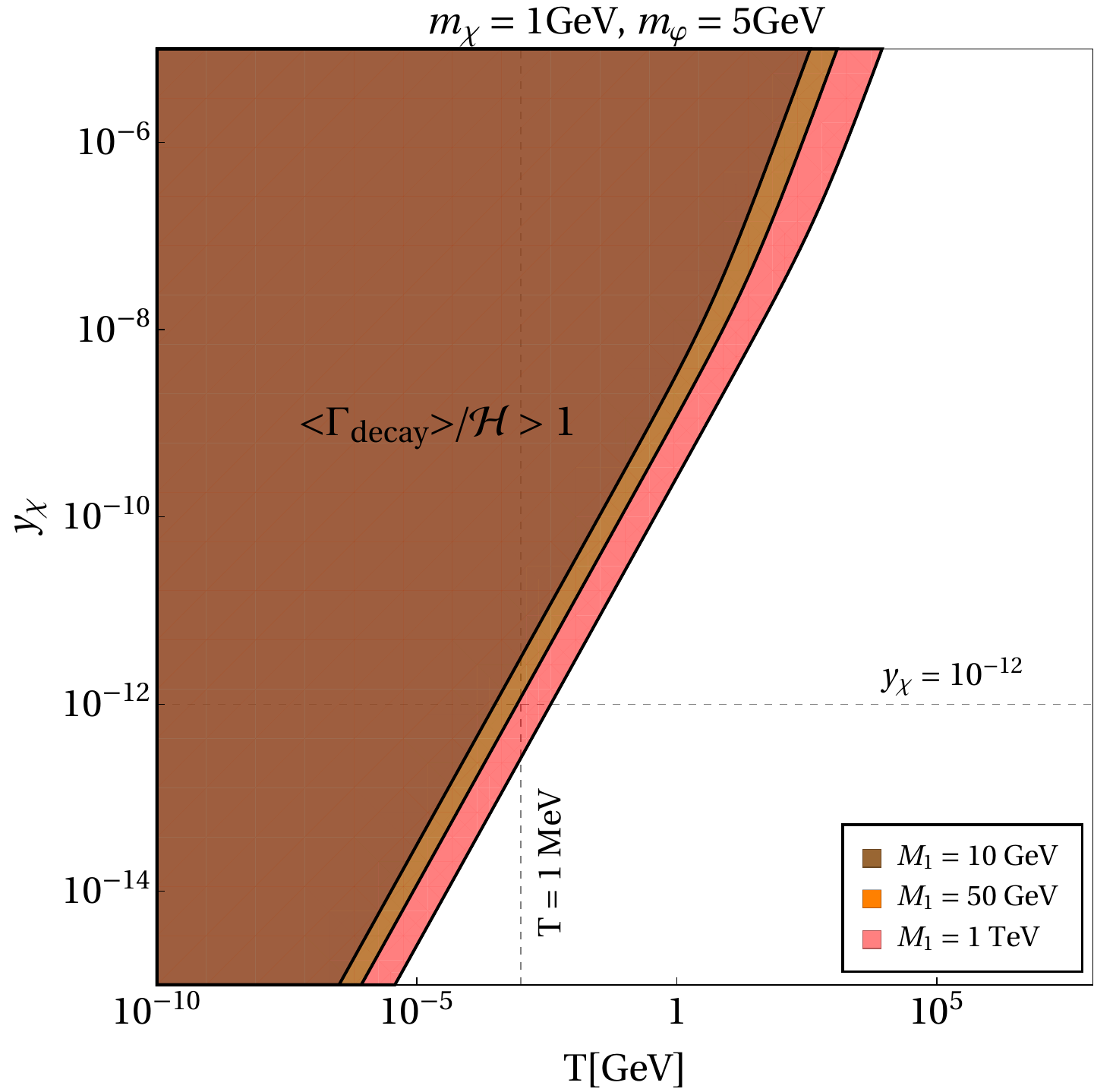}
    \includegraphics[width=0.495\textwidth]{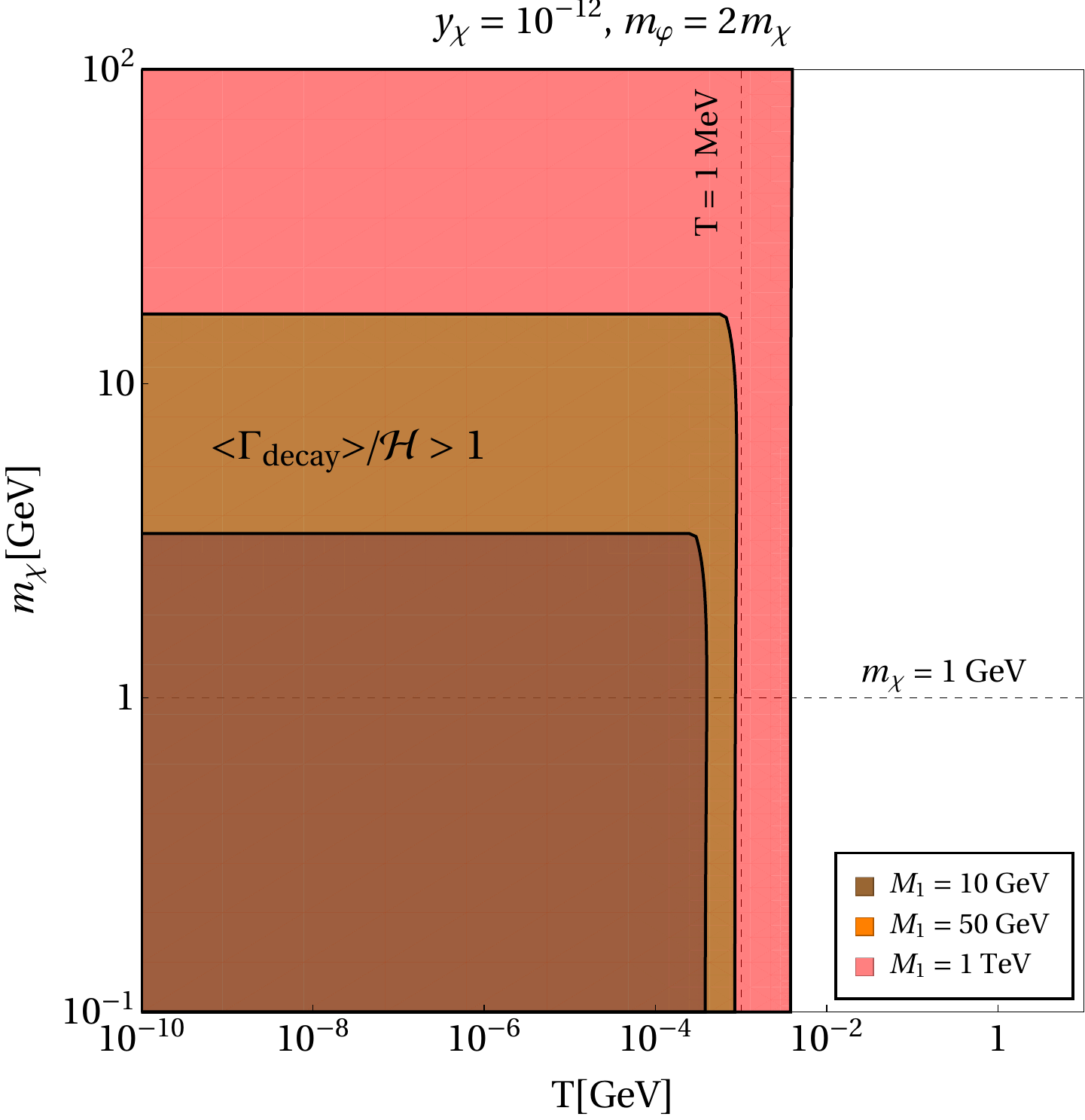}\\[10pt]
    \includegraphics[width=0.495\textwidth]{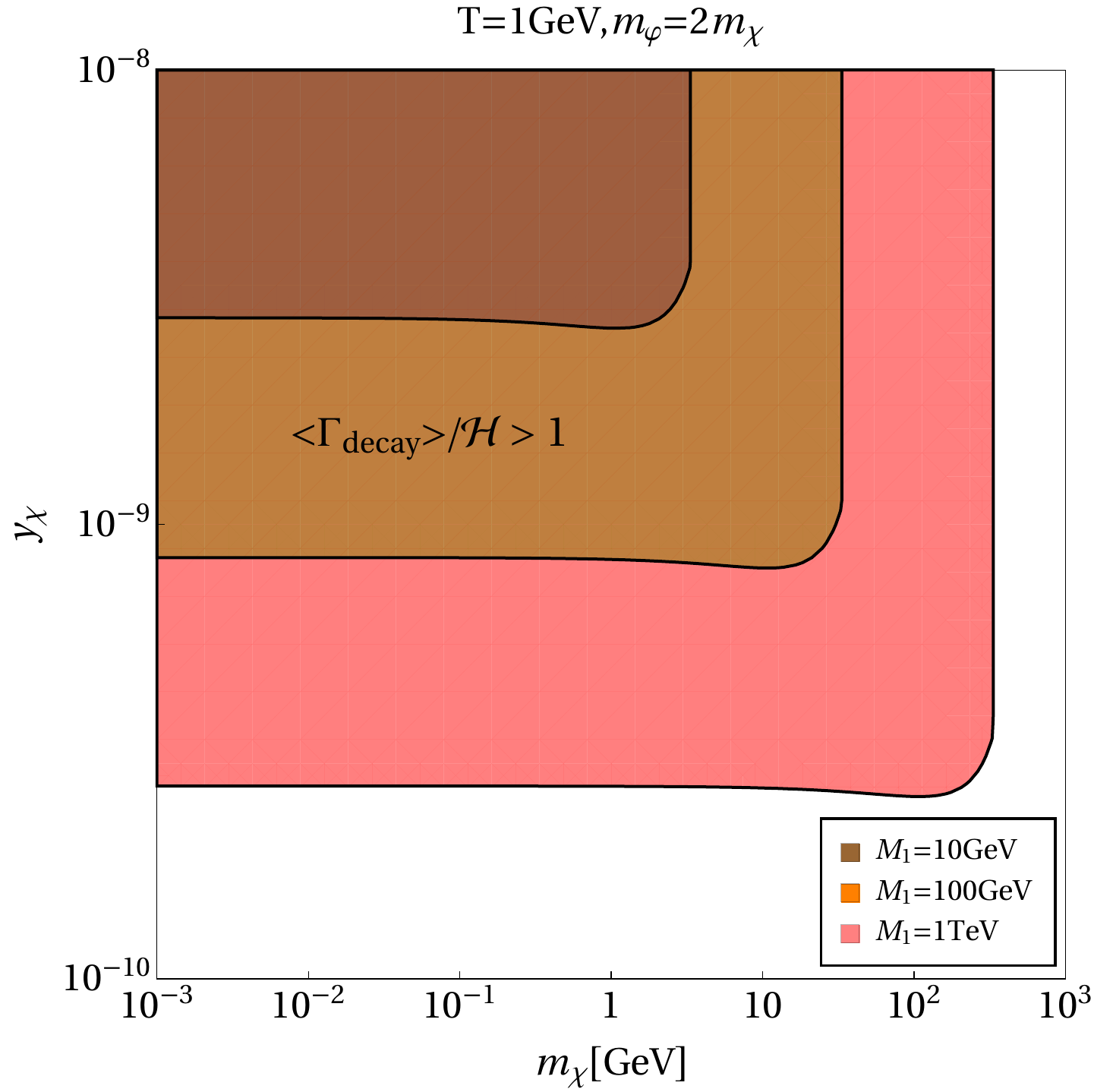}
    \caption{{\it Top Left:} Comparison of the thermally-averaged decay rate $\langle\Gamma_{N\to\varphi\,\chi}\rangle$ with the Hubble expansion rate $\mathcal{H}$ in the $(T,y_\chi)$ plane for fixed $m_{\chi\,,\varphi}$. {\it Top Right:} Decay rate in the $(T,m_\chi)$ plane for a fixed $y_\chi=10^{-12}$ and $m_\varphi=2\,m_\chi$. {\it Bottom:} Decay rate in the $(m_\chi,y_\chi)$ plane for a fixed $T=1$ GeV and $m_\varphi=2\,m_\chi$. The different curves are for different choices of $M_1$. In all cases, the shaded regions correspond to $\langle\Gamma_\text{decay}\rangle>\mathcal{H}$, where freeze-in is not applicable anymore.}
    \label{fig:rate}
\end{figure}

\begin{figure}[htb!]
    \centering
    \includegraphics[width=0.495\textwidth]{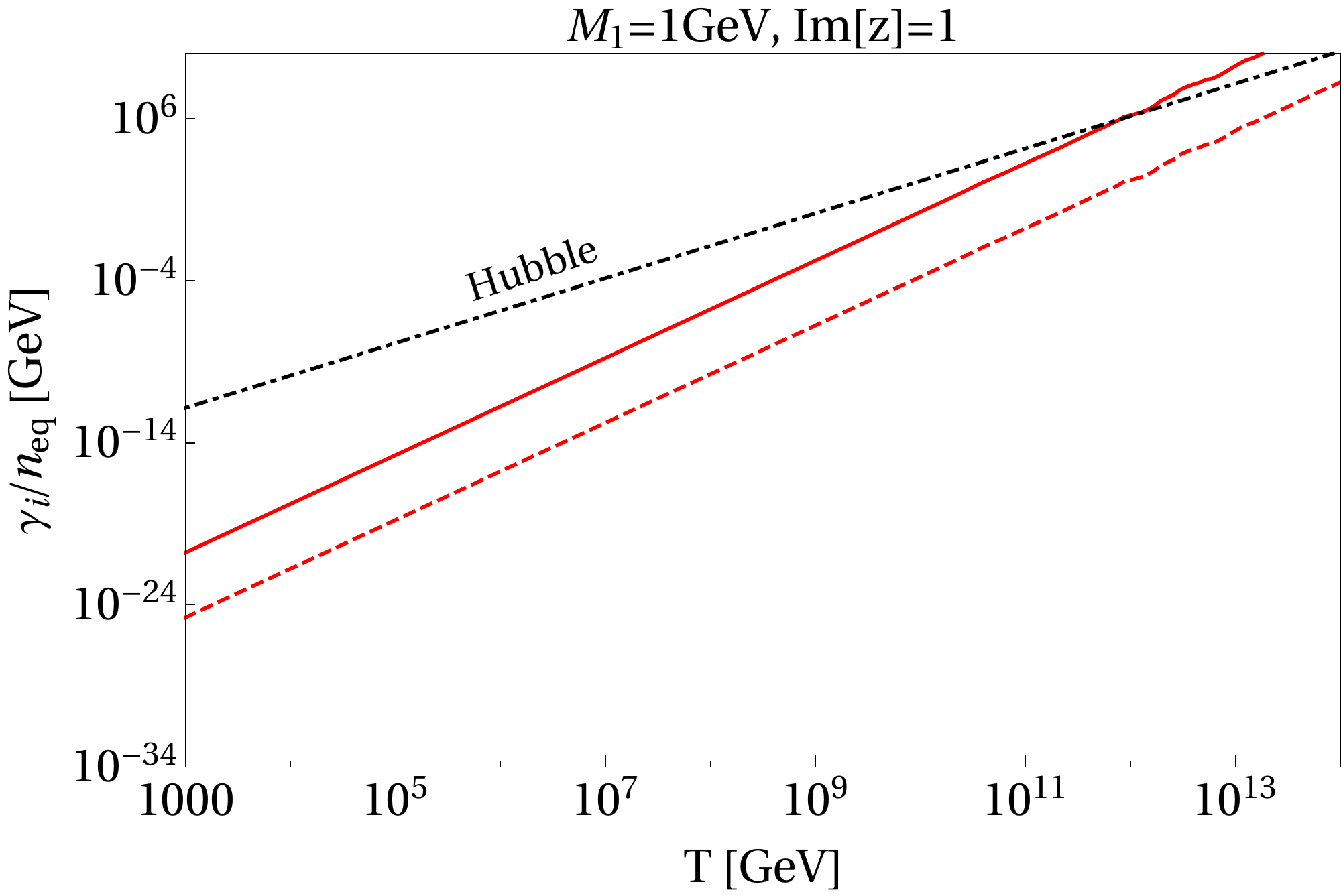}
    \includegraphics[width=0.495\textwidth]{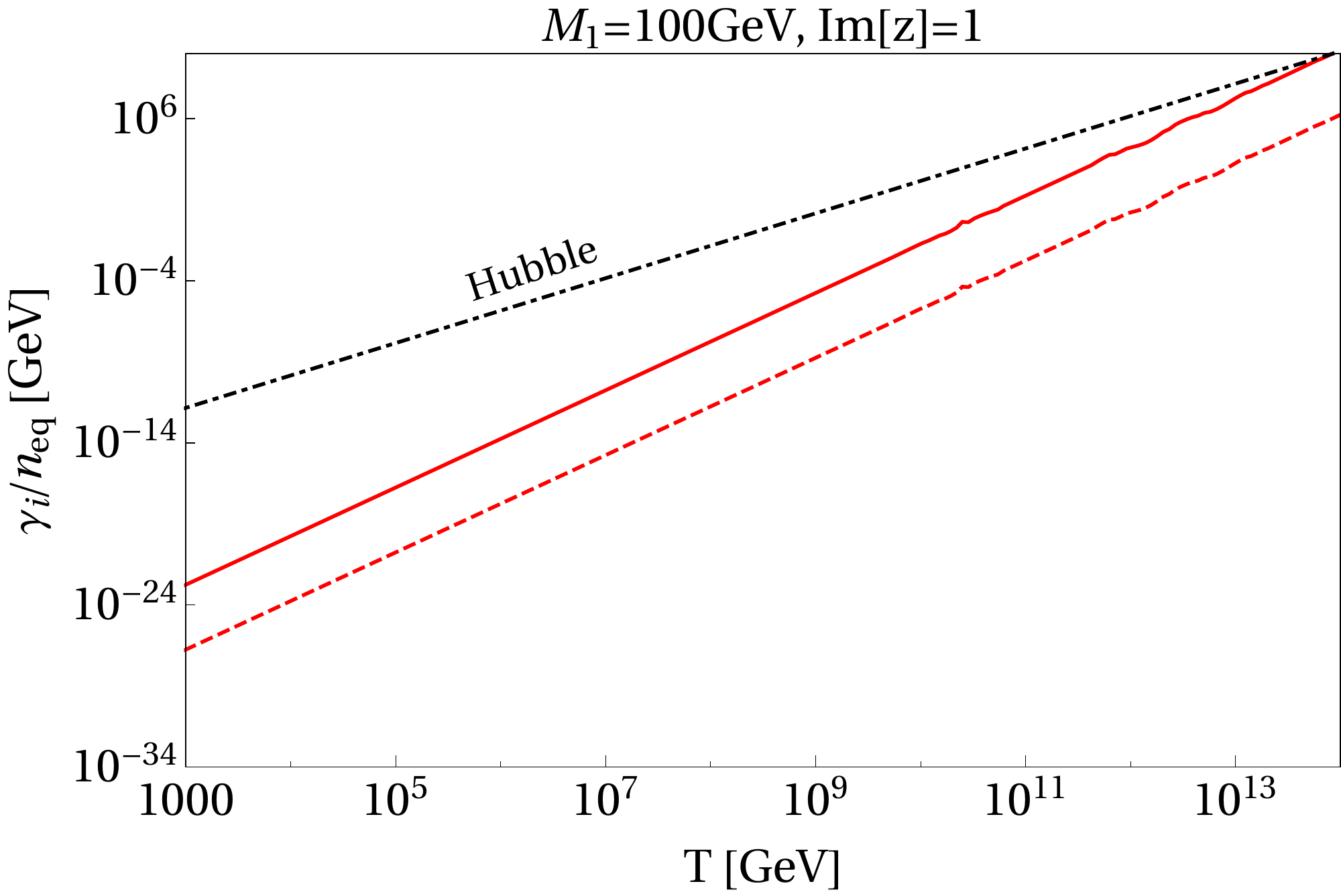}
    \caption{Comparison of 2-to-2 scattering rate (red) with the Hubble rate (black dot-dashed line), considering $m_\chi=50$ GeV, $m_\varphi=2\,m_\chi$, $M_1=1$ GeV (left) and $M_1=100$ GeV (right). The solid (dashed) red curves correspond to $y_\chi=10^{-5}\,(10^{-7})$.}
    \label{fig:22rate}
\end{figure}

\begin{figure}[htb!]
    \centering
    \includegraphics[width=0.495\textwidth]{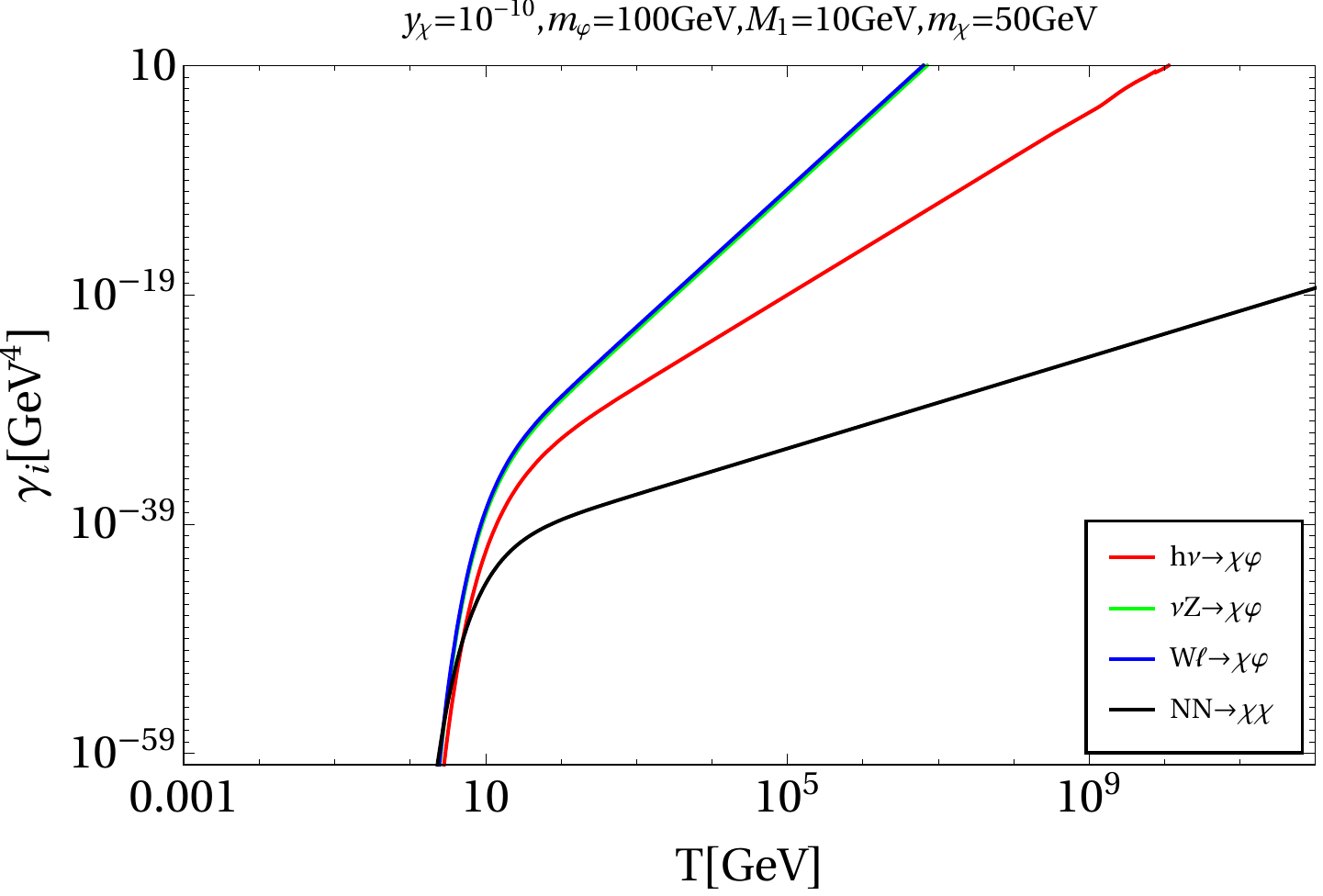} \includegraphics[width=0.495\textwidth]{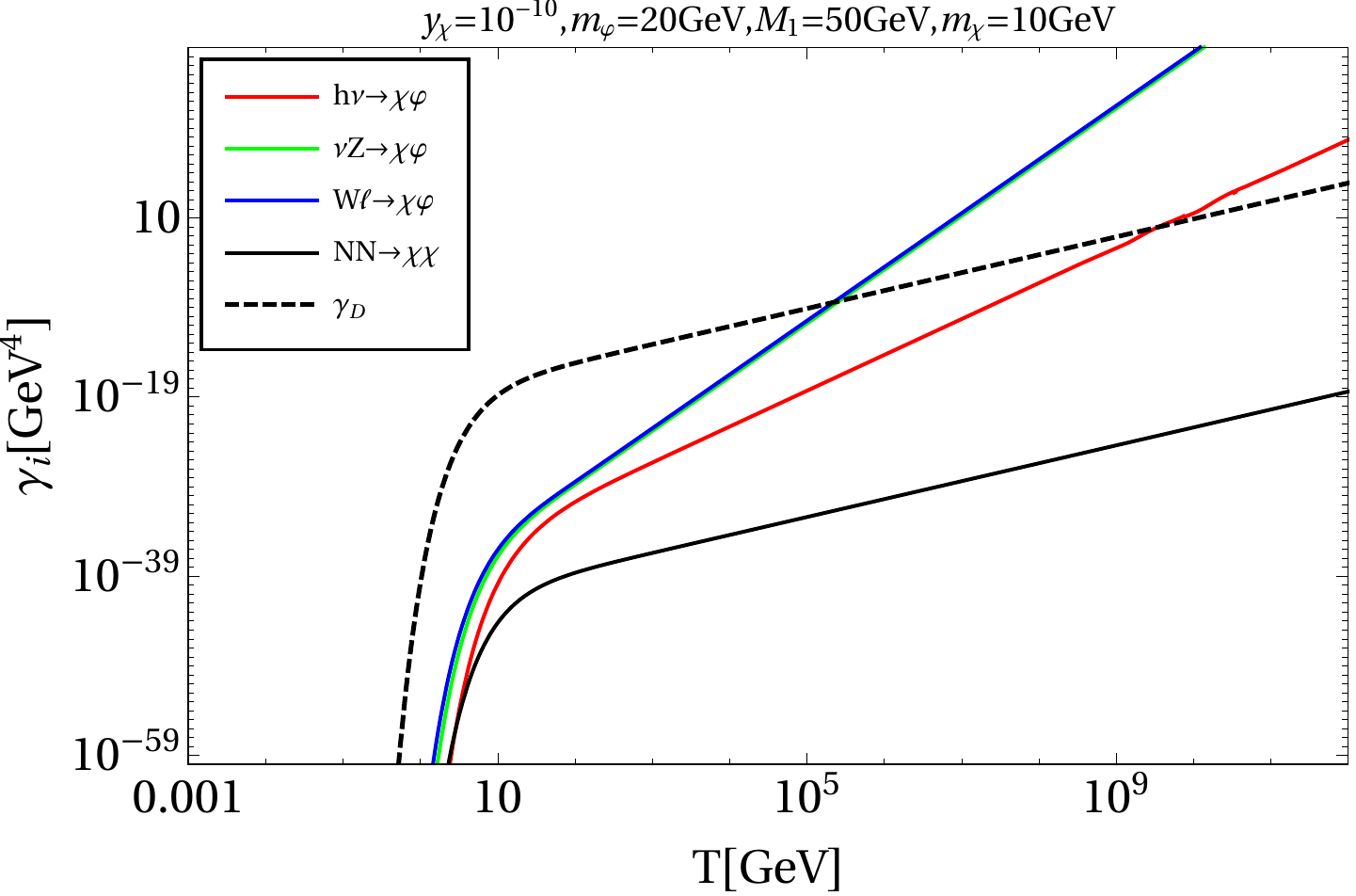}
    \caption{{\it Left:} Reaction densities for only 2-to-2 scattering channels. {\it Right:} Comparison of reaction densities due to scattering and decay (black, dashed curve).  All relevant parameters are kept fixed at the values mentioned in the plot labels.}
    \label{fig:gamma}
\end{figure}
\subsection{Comparison of the rates}
It is important to ensure that the DM does not thermalize with the SM bath in the early Universe, which is the primary requirement for freeze-in. Assuming that the $N\to\varphi\chi$ decay is kinematically available, we first find the region compatible with freeze-in from the requirement of $\langle\Gamma_{N\to\varphi\chi}\rangle<\mathcal{H}$. This is shown  in the top left panel of Fig.~\ref{fig:rate}, where we see the DM production can remain out-of-equilibrium till $T\sim\mathcal{O}(\text{MeV})$ (unshaded region)\footnote{Within the shaded region the DM can undergo thermal freeze-out, leading to a different production mechanism altogether.}. Here we have defined the thermally averaged decay width of RHN into DM as 
\begin{equation}\label{eq:therm-decay}
\langle\Gamma_{N\to\varphi\chi}\rangle = \Gamma_{N\to\varphi\chi}\,\frac{K_1\left(M_1/T\right)}{K_2\left(M_1/T\right)}\,,     
\end{equation}
where $\Gamma_{N\to\varphi\chi}\simeq y_\chi^2\,M_1/\left(8\,\pi\right)\,;M_1\gg m_{\varphi\,,\chi}$ is the decay width of $N_1$ into DM (see Appendix~\ref{sec:app-RHN-decay} for full expression) and $K_{1,2}$ are modified Bessel functions. This condition, in turn, puts a constraint on the DM-RHN coupling $y_\chi$. Note that, for lighter $M_1$ the DM remains out-of-equilibrium for a longer period of time before it equilibrates, since in that case the decay width becomes comparatively smaller, making the decay lifetime longer. In order for the freeze-in production via decay to stay non-thermal till $T\simeq 1$ GeV for $M_1=1$ TeV, one needs a very small $y_\chi\lesssim 10^{-10}$. The out-of-equilibrium condition is independent of the choice of $M_1$ till $T\sim 10$ MeV for a fixed $y_\chi=10^{-12}$, as we notice from the top right panel. One can not ensure the non-thermal production of the DM (via decay) below this point, irrespective of the choice of the masses $M_1\,,m_\chi$. The dashed vertical lines in the top panel show that for $m_\chi=1$ GeV and $M_1=10$ GeV, the non-thermal production is guaranteed for $T\gtrsim 1$ MeV, with $y_\chi=10^{-12}$. This constraint, however, becomes more stringent for a heavier RHN, as that results in a larger decay rate following Eq.~\eqref{eq:therm-decay}. For $M_1=1$ TeV, non-thermal production is valid for $T\gtrsim 100$ MeV with $\ydm\lesssim 10^{-10}$, as one can see from the top right panel. Finally, in the bottom panel we depict the allowed region of the parameter space in $(m_\chi,y_\chi)$ plane for different choices of $M_1$. We again note that the condition for freeze-in production is satisfied even at $T=1$ GeV for $y_\chi\lesssim 10^{-10}$. Since we are only considering $M_1<1$ TeV for phenomenological purposes, we choose the lower bound on temperature to be $T\gtrsim 100$ MeV to ensure non-thermal production from decay for $\ydm\lesssim 10^{-10}$. Needless to mention, to ensure $\Gamma_\text{decay}<\mathcal{H}$ at temperatures even below 1 GeV, one has to choose $y_\chi$ even smaller. For $M_1$ of mass $\sim$ MeV, this upper limit on $y_\chi$ can be relaxed since in that case the decay width itself is small enough to keep the DM out of equilibrium. Considering a conservative limit, we will thus keep ourselves confined to $y_\chi\lesssim 10^{-10}$ in computing the DM yield via decay.   

For DM production via scattering, we compare the rates of 2-to-2 scattering processes with the Hubble rate. The reaction rate for scattering process is given by $\mathcal{R}=n_\text{eq}\,\langle\sigma v\rangle\equiv\gamma_\text{ann}/n_\text{eq}$, where we consider $n_\text{eq}$ to be the equilibrium number density of the SM particles in the initial state. In this case the bound on $y_\chi$ can be significantly relaxed, as one can see from Fig.~\ref{fig:22rate}. Note that, for $y_\chi=10^{-7}$ the processes can safely be considered to be out of equilibrium even at high temperatures. This is expected since the thermal averaged production cross-section $\langle\sigma v\rangle\propto\left|\pmns\,U_{\nu N}\right|^2\,y_\chi^2$ in this case. In the high temperature regime $(T\gtrsim 1\,\rm TeV)$ when the electroweak symmetry is exact, all the SM particles can be assumed to be massless. We would like to mention here that although the 2-to-2 scattering rate depends on the DM mass, but with $\ydm\lesssim 10^{-7}$ we stay safely below the Hubble rate for $T\gtrsim 10^3$ GeV. For the scattering channels, therefore, a conservative limit on $y_\chi\lesssim 10^{-7}$, which is about three orders of magnitude relaxed than that considered for the case of decay.

In the presence of DM production both via decay and scattering, the contribution from decay generally wins in the lower temperature region, where IR freeze-in becomes important. In order to establish that, in Fig.~\ref{fig:gamma} we show the reaction densities $\gamma_i$ as a function of bath temperature $T$ for the case when {\it only} scattering is allowed (in the left panel), i.e., $M_1<m_\chi+m_\varphi$, and where both scatterings and decays are allowed (in the right panel). Here we find, in presence of both, the decay indeed dominates in the low temperature regime. Since we are typically considering the IR freeze-in scenario, where the DM yield becomes important at later times (low temperatures), hence in case where the decay channels are open, we can ignore the 2-to-2 DM production channels. Therefore, our analysis will be divided into two categories: (a) $M_1>m_\chi+m_\varphi$ for which the DM production from decay is dominant, and (b) $M_1\leq m_\chi+m_\varphi$ for which DM production from scattering is important (the decay is kinematically forbidden). Before moving on we would like to mention that the present model provides four free parameters (we fix the Higgs portal coupling $\lambda_{H\varphi}$ as we will elaborate later): $\{M_1\,,m_\chi\,,m_\varphi\,,y_\chi\}\,,$ which determine the viable parameter space for the DM. 

\subsection{DM production via RHN decay}
\label{sec:decay}
For $M_1>m_\chi+m_\varphi$, the DM production channel from RHN decay is kinematically available. The asymptotic DM yield can be analytically computed by integrating Eq.~\eqref{eq:beq} 
\begin{equation}
Y_\chi(T=0)\approx \frac{405\,g_N}{4\,\pi^2\,g_{\star s}\,\sqrt{g_{\star\rho}}}\,\sqrt{\frac{5}{2}}\,\frac{M_P\,\Gamma_{N_1\to\chi\varphi}}{M_1^2}\,, 
\label{eq:yld-decay}
\end{equation}
where $g_N=2$ is the number of degrees of freedom for RHN. Here $g_{\star s}$ and $g_{\star\rho}$ are the effective number of relativistic degrees of freedom contributing to the entropy and energy density respectively, while $M_P$ is the reduced Planck mass. The relic abundance at present epoch $T=T_0$ can then be obtained using
\begin{equation}
\Omega_\chi h^2 = \left(2.75\times 10^8\right) \left(\frac{\mdm}{\text{GeV}}\right) Y_\chi(T_0)\,,
\label{eq:relicX}    
\end{equation}
which needs to satisfy the value as measured by Planck: $\Omega_\text{DM} h^2 = 0.11933\pm 0.00091$~\cite{Planck:2018vyg}. We find that right relic density can be obtained for $y_\chi\simeq 7\times 10^{-12}$ for a DM mass of $m_\chi=1$ GeV with $M_1=10$ GeV and $m_\varphi=2\,m_\chi$. Note that the size of the coupling $y_\chi$ typically falls within the ballpark where non-thermal DM production is viable [cf. Fig.~\ref{fig:rate}].
\begin{figure}[htb!]
    \centering
    \includegraphics[width=0.49\textwidth]{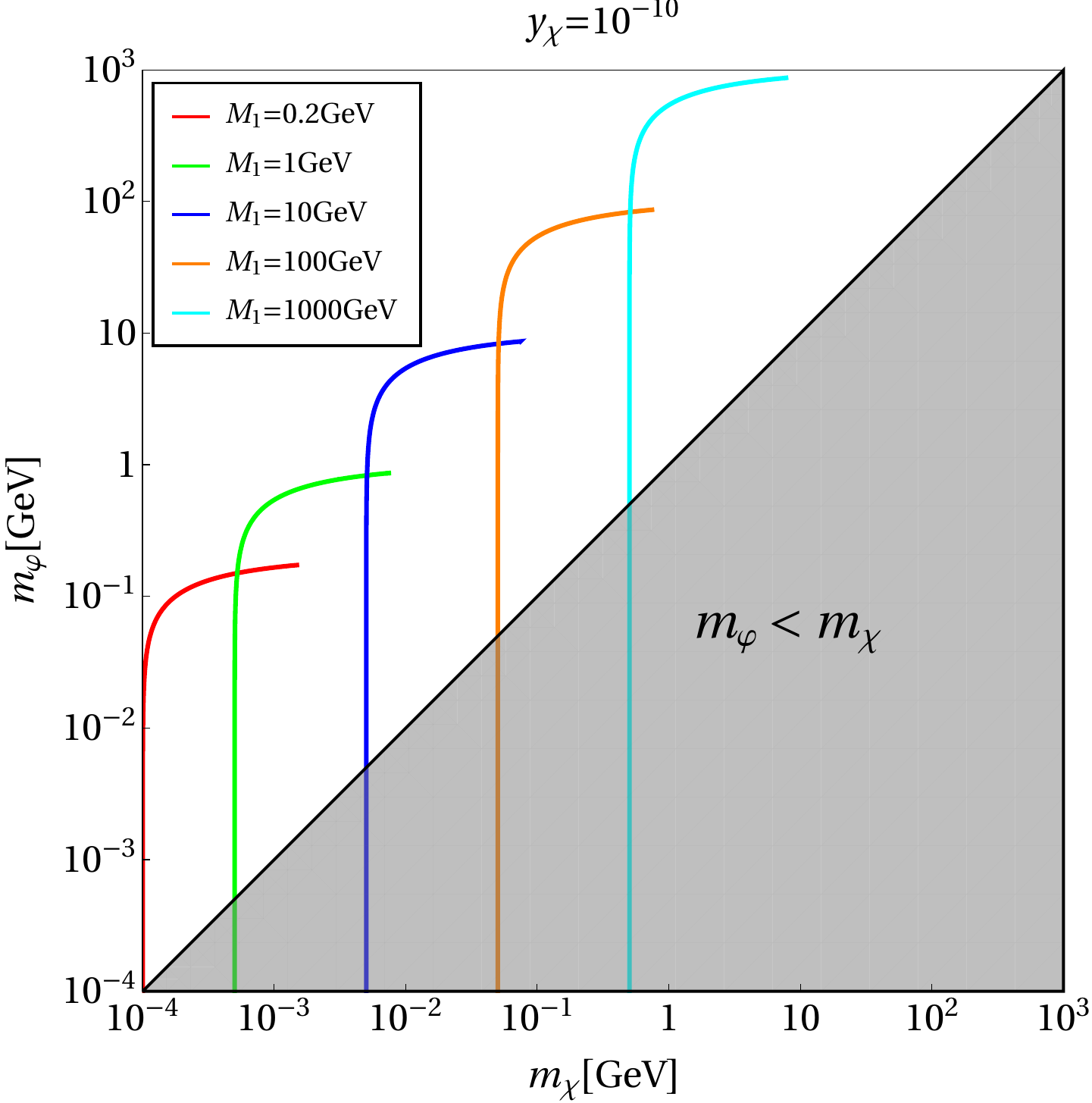} \includegraphics[width=0.49\textwidth]{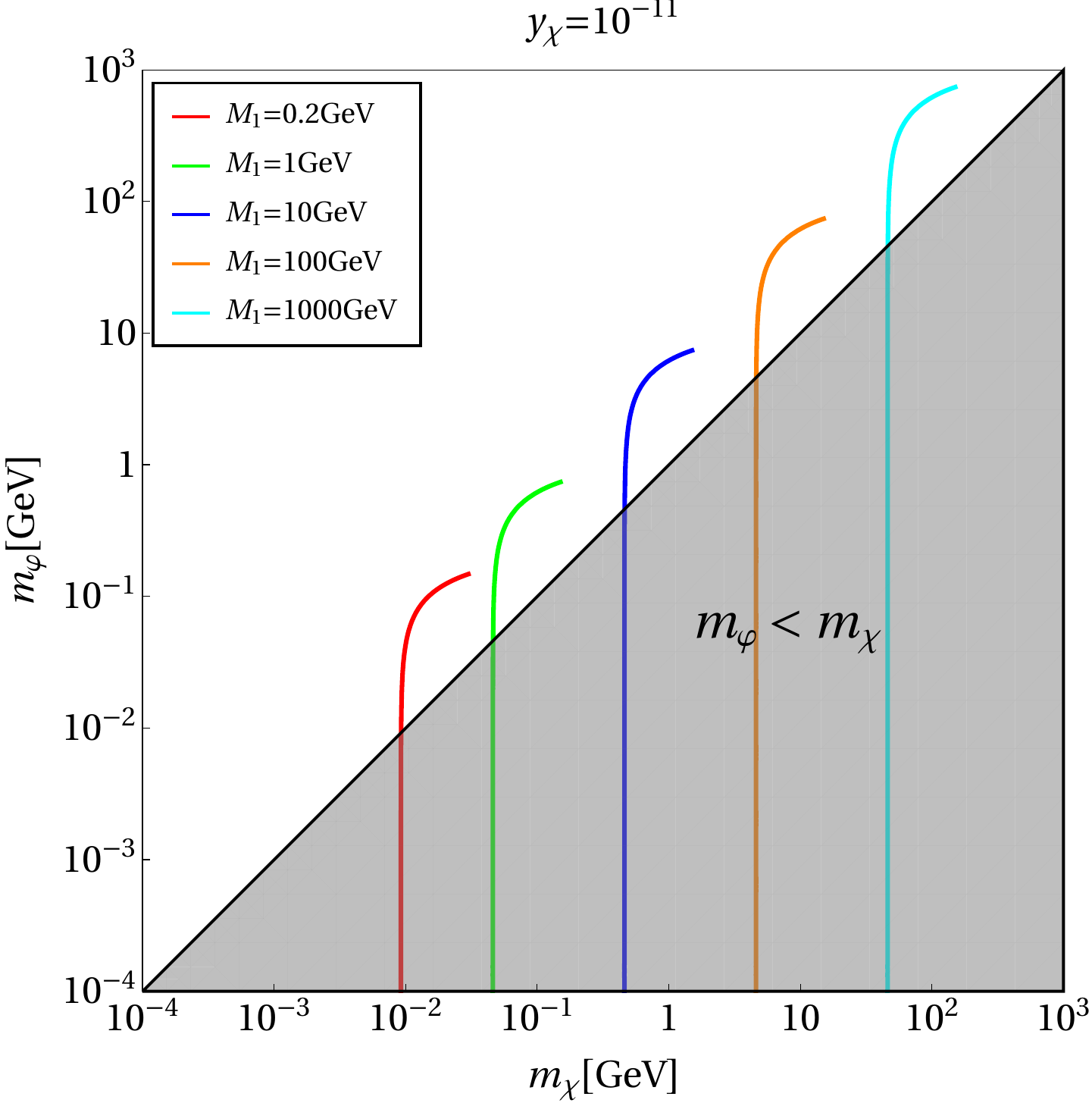} \\[10pt]
    \includegraphics[width=0.49\textwidth]{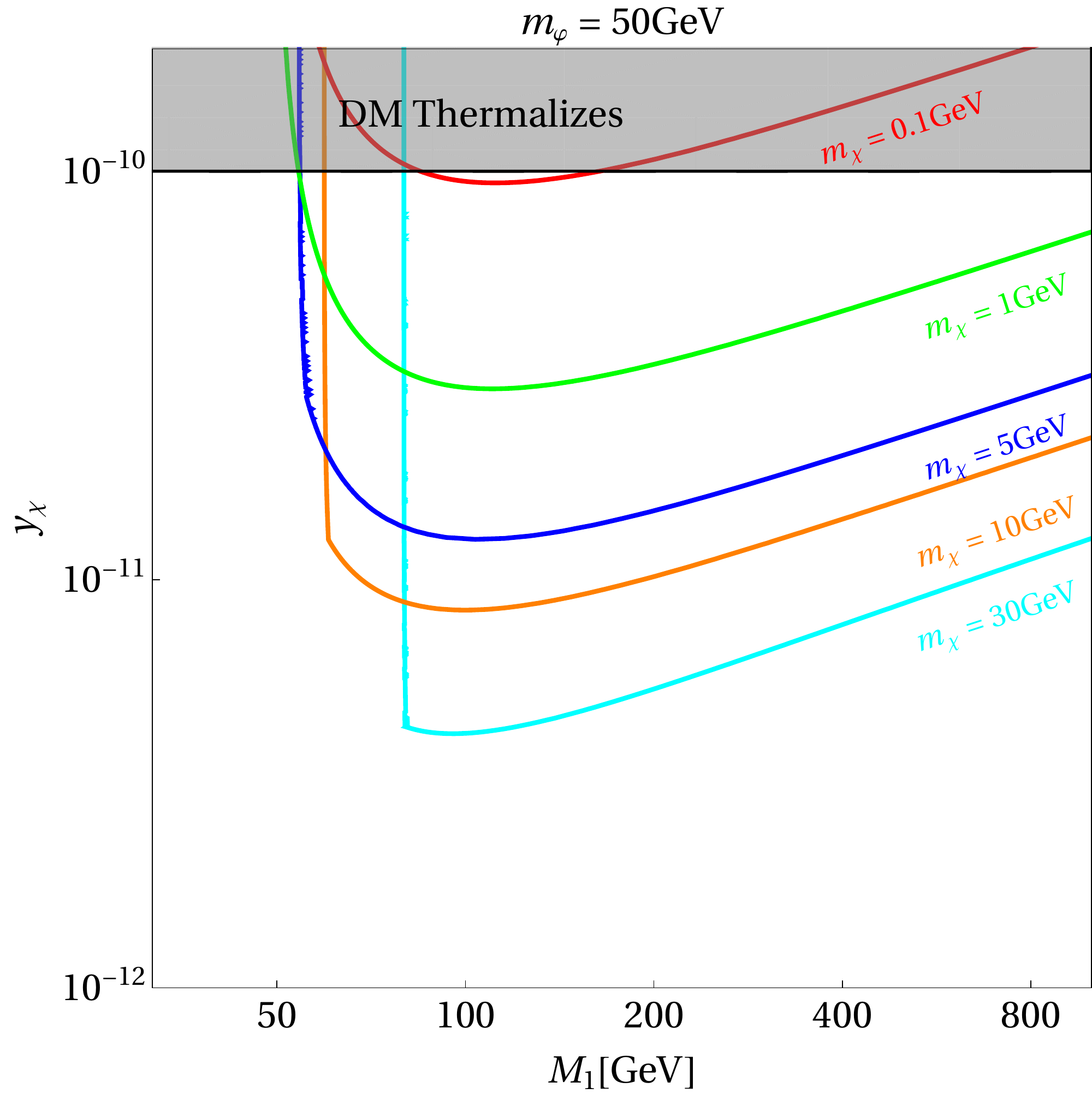}
    \includegraphics[width=0.49\textwidth]{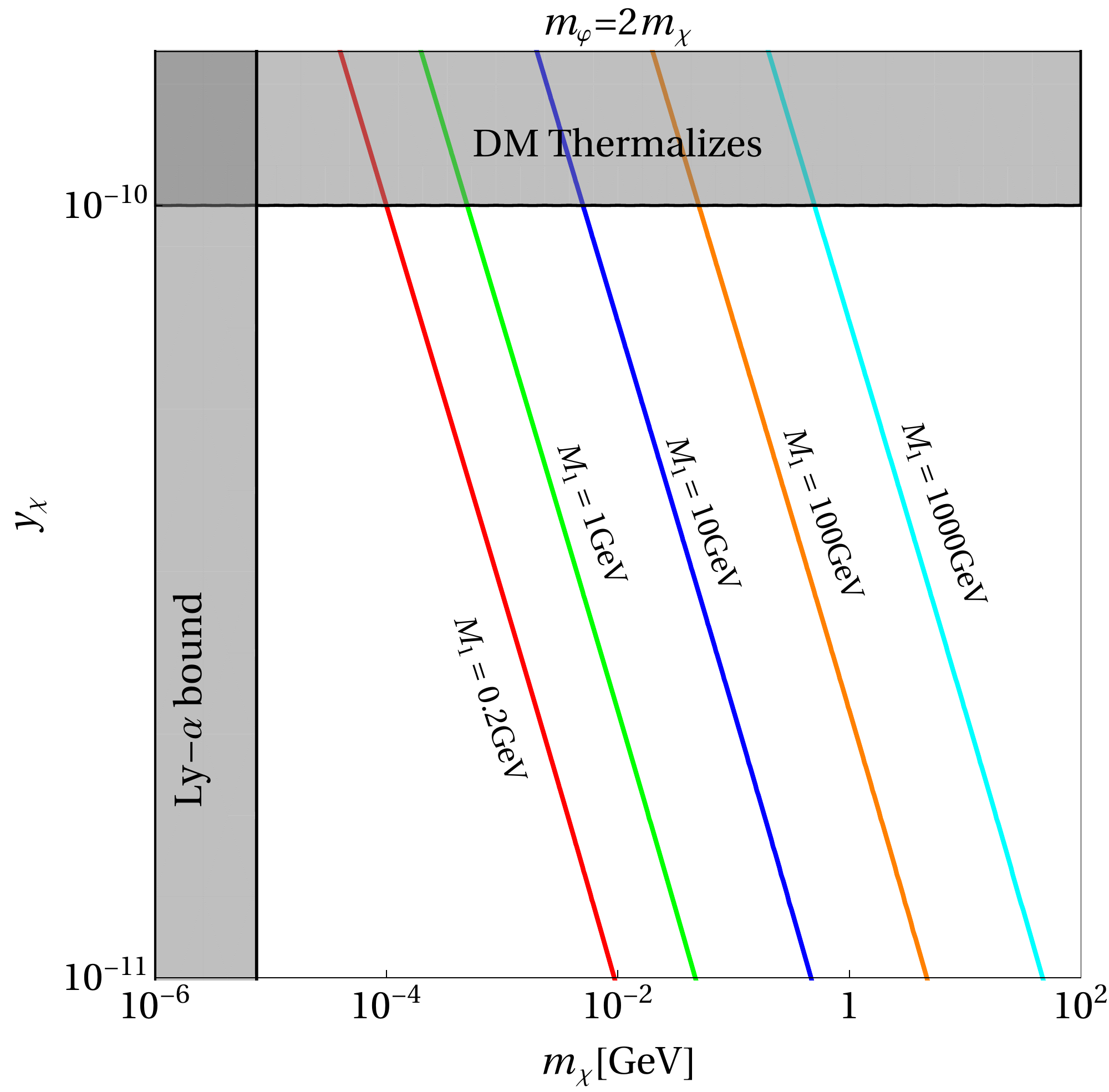}
    \caption{{\it Top Left:} Contours of different colors satisfy the central value of the observed relic density for different choices of RHN masses for $y_\chi=10^{-10}$. {\it Top Right:} For $y_\chi=10^{-11}$. In both plots, the gray-shaded region is for $m_\varphi<m_\chi$ in which case $\chi$ is no longer the DM. {\it Bottom Left:} Relic density contours for different choices of DM masses for $m_\varphi=50$ GeV in $(M_1,y_\chi)$ plane. {\it Bottom Right:} Different contours correspond to right relic abundance for different choices of $M_1$ for $m_\varphi=2\,m_\chi$. The gray-shaded regions are forbidden from DM thermalization condition via decay and the Lyman-$\alpha$ bound (see text).}
    \label{fig:decay-rel-cntr}
\end{figure}
The contours of the observed relic density with different choices of the masses of the decaying RHN are shown in the top left panel of Fig.~\ref{fig:decay-rel-cntr} for a fixed $y_\chi$ in $(m_\chi,m_\varphi)$ plane. For a fixed $M_1$ the contours stop at a particular DM mass since beyond that the decay is kinematically disallowed. With increase in $M_1$ it is possible to obtain the correct relic density for larger $m_\chi$ for a fixed $y_\chi$ since $Y_\chi\propto 1/M_1$ as one can see from Eq.~\eqref{eq:yld-decay} (on using the decay width from Appendix~\ref{sec:app-RHN-decay}). On decreasing the DM Yukawa coupling $y_\chi$, for a fixed $M_1$, one expects to obtain the right abundance for a larger DM mass. This is reflected in the top right panel, where we have chosen $y_\chi=10^{-11}$. A part of the parameter space becomes forbidden where $m_\varphi<m_\chi$, making the scalar $\varphi$ as the lightest dark sector state with odd ${Z}_2$ symmetry. In the bottom panel of Fig.~\ref{fig:decay-rel-cntr} we show contours satisfying relic abundance in $(M_1,y_\chi)$ plane (left) for fixed $m_\varphi=50$ GeV and $(m_\chi,y_\chi)$ plane (right) for $m_\varphi=2\,m_\chi$. In $(M_1,y_\chi)$ plane we show contours corresponding to some fixed choices of the DM mass $m_\chi$. We see that a larger DM mass requires a smaller $y_\chi$ to produce the required abundance, as already seen from the top panel. For $y_\chi\gtrsim 10^{-10}$ the DM can thermalize in the early Universe as mentioned earlier and freeze-in is no longer viable. The kinematical cut-off in each contour, as shown by the lines parallel to the vertical axis, corresponds to the value of $M_1$ below which $N_1$ decay is not possible. Finally, the bottom right panel summarizes the allowed DM parameter space in $(m_\chi,y_\chi)$ plane for a fixed $m_\varphi$, where a larger $M_1$ produces observed relic for a larger coupling. It is important to note here that for $M_1\lesssim\mathcal{O}$(MeV), the DM mass has to be extremely light for larger $y_\chi$ to satisfy the relic abundance. However, this is constrained from the measurements of the free-streaming of warm DM (WDM) from Lyman-$\alpha$ flux-power spectra~\cite{Baur:2015jsy, Irsic:2017ixq, Ballesteros:2020adh, DEramo:2020gpr} that only allows DM mass $\gtrsim 7.5$ keV. This is depicted as the Lyman-$\alpha$ bound. This bound is slightly stronger than the theoretical bound on fermionic DM from phase-space considerations -- the so-called Tremaine-Gunn bound~\cite{Tremaine:1979we}. For smaller $y_\chi$, this constraint on the parameter space does not exist anymore since in that case one has to go to a heavier DM to satisfy the freeze-in relic density. It is important to note here that RHN masses below 1 GeV will also be constrained by BBN~\cite{Dolgov:2000jw, Boyarsky:2020dzc, Sabti:2020yrt} and CMB~\cite{Vincent:2014rja, Mastrototaro:2021wzl} if, depending on the active-sterile neutrino mixing, their lifetime is longer than about $0.1$ sec.  However, since we are only dealing with the DM coupling $\ydm$ in this plot, the cosmological constraints on RHN are not shown here, which can always be satisfied with the appropriate choice of the active-sterile mixing.
\begin{figure}[t!]
    \centering
    \includegraphics[scale=0.36]{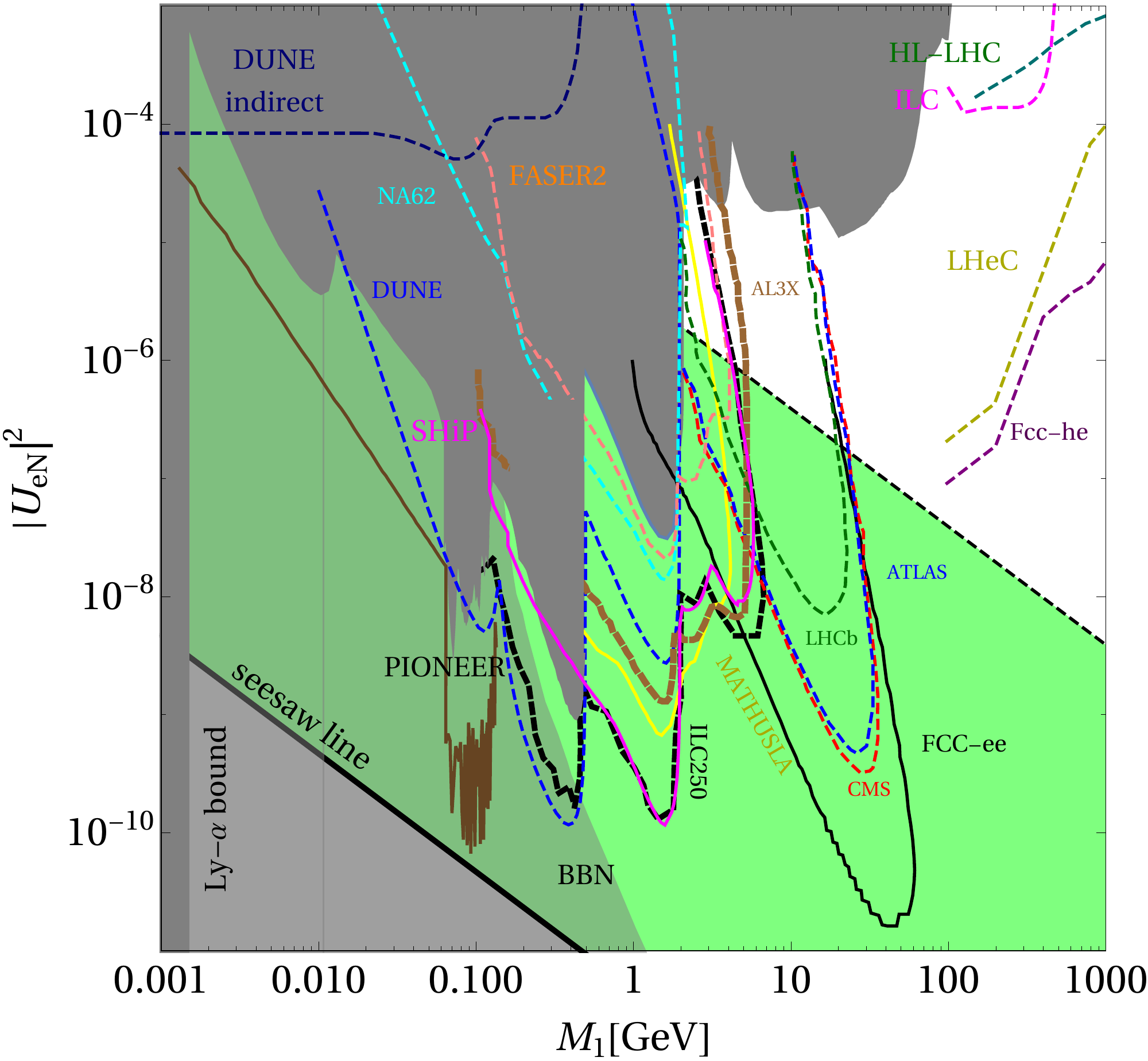}~~~
    \includegraphics[scale=0.36]{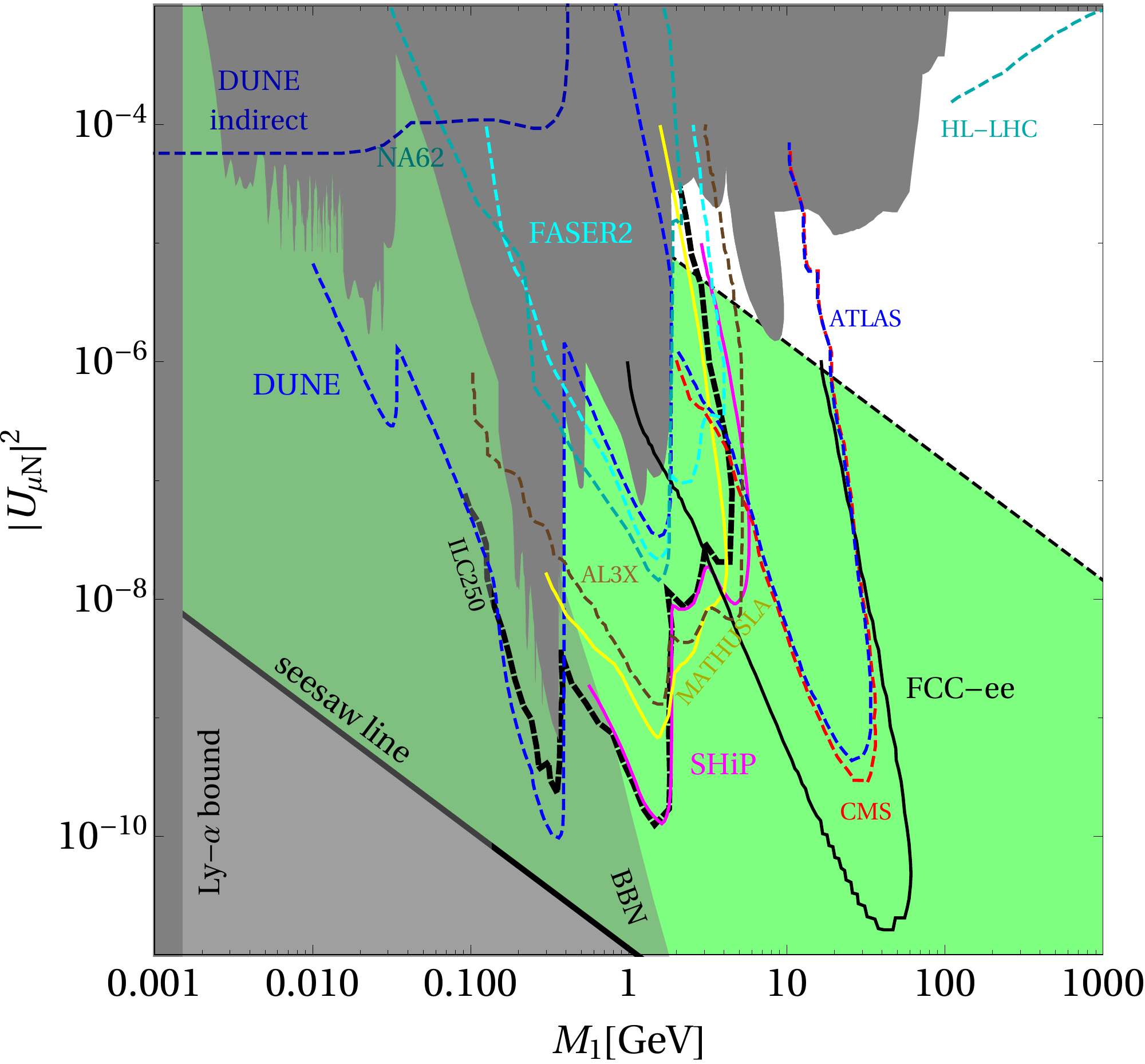}
    \\[10pt]
    \includegraphics[scale=0.36]{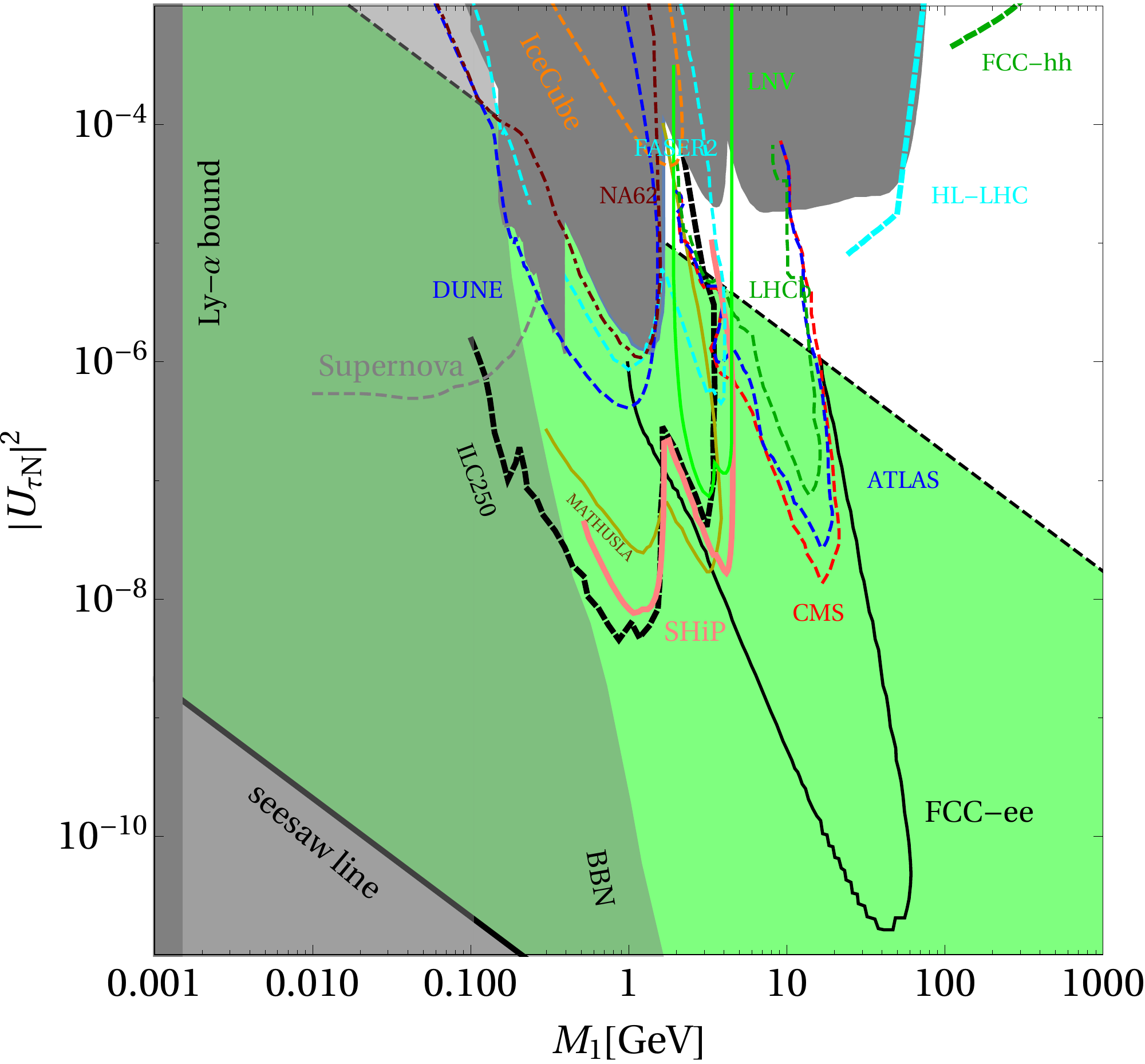}
    \caption{FIMP DM region superimposed on the RHN parameter space in the plane of RHN mass and its coupling to the electron, muon and tau flavors. In each panel, the green-shaded region corresponds to the observed DM abundance in the RHN parameter space for normal hierarchy of neutrino masses. Here we have fixed $y_{\chi}=10^{-10}$ to ensure non-thermal production of DM. The black-dashed diagonal lines correspond to the seesaw lines for $\text{Im}[z]=0$ (lower) and $\text{Im}[z]=7$ (upper). The gray-shaded regions are excluded by various RHN constraints. The future RHN sensitivities are also shown for comparison. See text for details. }
    \label{fig:expt-decay}
\end{figure}

To obtain the net parameter space for the DM satisfying relic abundance we scan over the DM and RHN masses, while keeping $m_\varphi=2\,m_\chi$ and fixing the DM Yukawa coupling $y_\chi=10^{-10}$, ensuring the DM production via decay remains non-thermal as discussed earlier. We also fix  $\text{Re}[z]=0.1$, and vary $\text{Im}[z]\in [0.1,7.0]$ in the CI parametrization [cf.~Eq.~\eqref{eq:CI}] for the Dirac Yukawa coupling. Note that $\text{Im}[z]=0$ would be the canonical seesaw case with very small active-sterile neutrino mixing, which is shown by the lower black dashed line in Fig.~\ref{fig:expt-decay}, whereas the upper black dashed line corresponds to $\text{Im}[z]=7$ which is the rough upper limit on the mixing from charged lepton flavor violation~\cite{Dev:2014qbx}.  Since the DM coupling $y_\chi$ that controls the relic abundance is uncorrelated with the RHN Yukawa coupling for the decay scenario (except for the tiny change in the RHN decay width), hence for a given choice of $y_\chi$ satisfying the relic density, it is always possible to have a viable parameter space in the RHN mass-mixing plane for the DM, which also satisfies the light neutrino mass constraints. This is shown by the green shaded regions in Fig.~\ref{fig:expt-decay}: the upper left panel for electron mixing (here $U_{eN}$ stands for $(U_{\nu N})_{e1}$), the upper right panel for muon mixing and the bottom panel for tau mixing. The choice of $y_\chi$ is motivated from the bottom left panel of Fig.~\ref{fig:decay-rel-cntr} to ensure that the DM production takes place via freeze-in. 

Also shown in Fig.~\ref{fig:expt-decay} are the current RHN exclusion regions (gray shaded) from various cosmological observations (such as BBN, CMB, Lyman-$\alpha$), as well as laboratory constraints (such as beta decays, meson decays, beam-dump searches, precision electroweak tests, and direct collider searches); see Refs.~\cite{Bolton:2019pcu,Boyarsky:2020dzc,Bondarenko:2021cpc,Boiarska:2021yho, Bolton:2022pyf,Barouki:2022bkt, Nojiri:2022xqn} for details.\footnote{The data used for plotting the constraints is available on the website \url{www.sterile-neutrino.org}.}The future RHN sensitivities are also shown (unshaded curves) for comparison. We see that part of the DM relic density allowed parameter space for $\text{Im}[z]\gtrsim 1$ lies within reach of future sensitivity of beam dump and collider experiments. Because of $M_N^{-1/2}$ dependence [cf., Eq.~\eqref{eq:umatrix}], a larger $M_1$ satisfies the light neutrino mass for comparatively lower $\left|U_{\nu N}\right|^2$, while the $\left|U_{\nu N}\right|^2$ coupling is boosted for higher $\text{Im}[z]$, improving the experimental reach. As we have already seen in Fig.~\ref{fig:decay-rel-cntr}, a larger DM Yukawa requires a lighter DM to produce the observed abundance, which is constrained from the WDM limit due to Lyman-$\alpha$ constraints. Here we have considered the conservative limit of 7.5 keV on fermion DM mass. This is shown by the black vertical line that forbids $M_1\lesssim 1$ MeV in the decay scenario [cf.~Fig.~\ref{fig:decay-rel-cntr}]. Our analysis presented here is for NH; we have repeated the same analysis for IH, however we do not show them here since there is no visible change in the resulting parameter space. On considering a smaller DM Yukawa coupling, the parameter space remains unchanged, except that the WDM bound does not apply anymore since the DM mass is always found to be above the keV scale to satisfy the relic density bound. 
\subsection{DM production via RHN scattering}
\label{sec:ann}
Next, let us take up the case where $M_1<m_\chi+m_\varphi$, making only the 2-to-2 processes available for DM production. Because of of both $s$ and $t$ channel contributions [cf.~Fig.~\ref{fig:DM-prod}], it is difficult to obtain an exact analytical solution of the BEs in this case. Nevertheless, an approximate analytical solution to the BEs for freeze-in production via scattering has been obtained in Ref.~\cite{Becker:2018rve} in the limit $M_N\ll m_\chi\approx m_\varphi$. As one can notice from Fig.~\ref{fig:gamma}, although the $t$-channel process has a negligible reaction rate compared to the $s$-channel process at high temperatures, but at lower temperature they become comparable. Since for IR freeze-in the DM yield at lower temperatures is important, hence we must include contributions from both channels in the present scenario. This can also be understood in the following way: The final DM yield and hence relic abundance has a $y_\chi^2\,Y_D^2$ dependence on the couplings for the $s$-channel process, while for $t$-channel process this dependence becomes $y_\chi^4$. This has also been elaborated in Ref.~\cite{Becker:2018rve}, where it was shown that the contribution of the heavy neutrino scattering processes account for $\sim 80\%$ of the produced DM in case of $\ydm\approx Y_D$. In our numerical analysis we take into account both $s$ and $t$-channel-mediated processes without any prejudice, since $Y_D$ can vary over a large range from about $10^{-8}$ to $10^{-2}$, depending on the values of ${\rm Im}[z]$ and $M_1$. 

\begin{figure}[t!]
    \centering
    \includegraphics[width=0.495\textwidth]{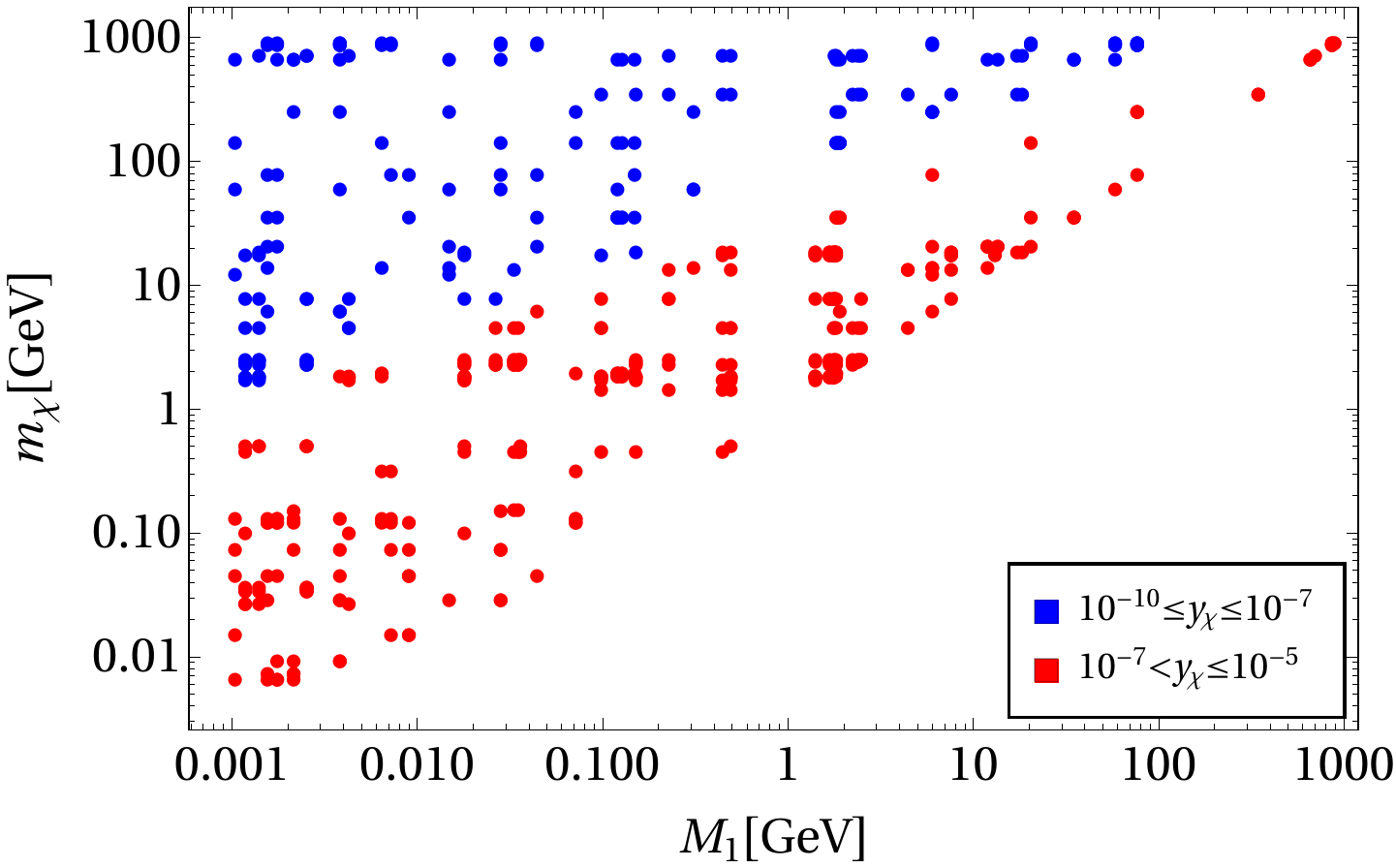} \includegraphics[width=0.495\textwidth]{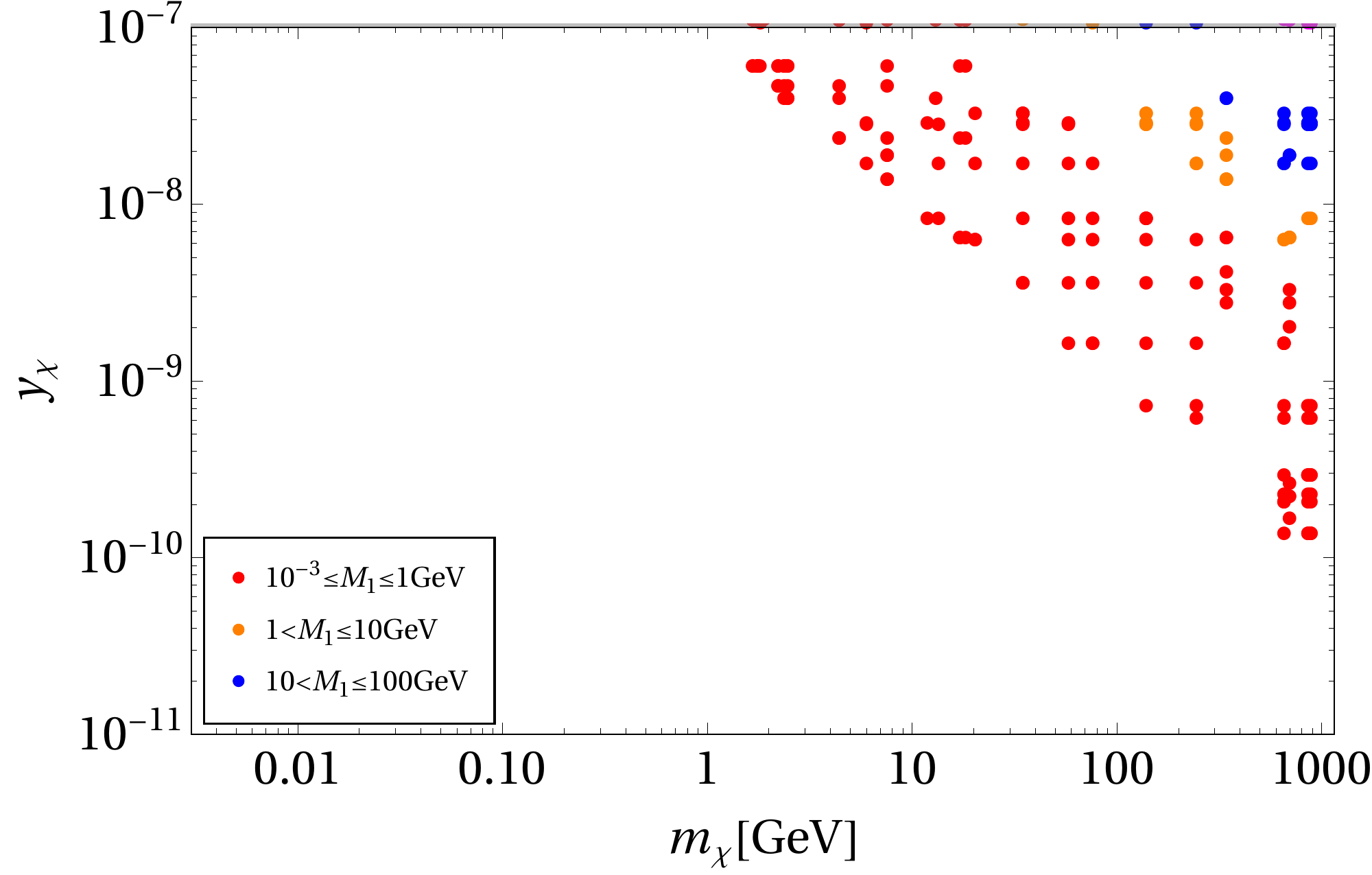} \\[10pt]
    \includegraphics[width=0.495\textwidth]{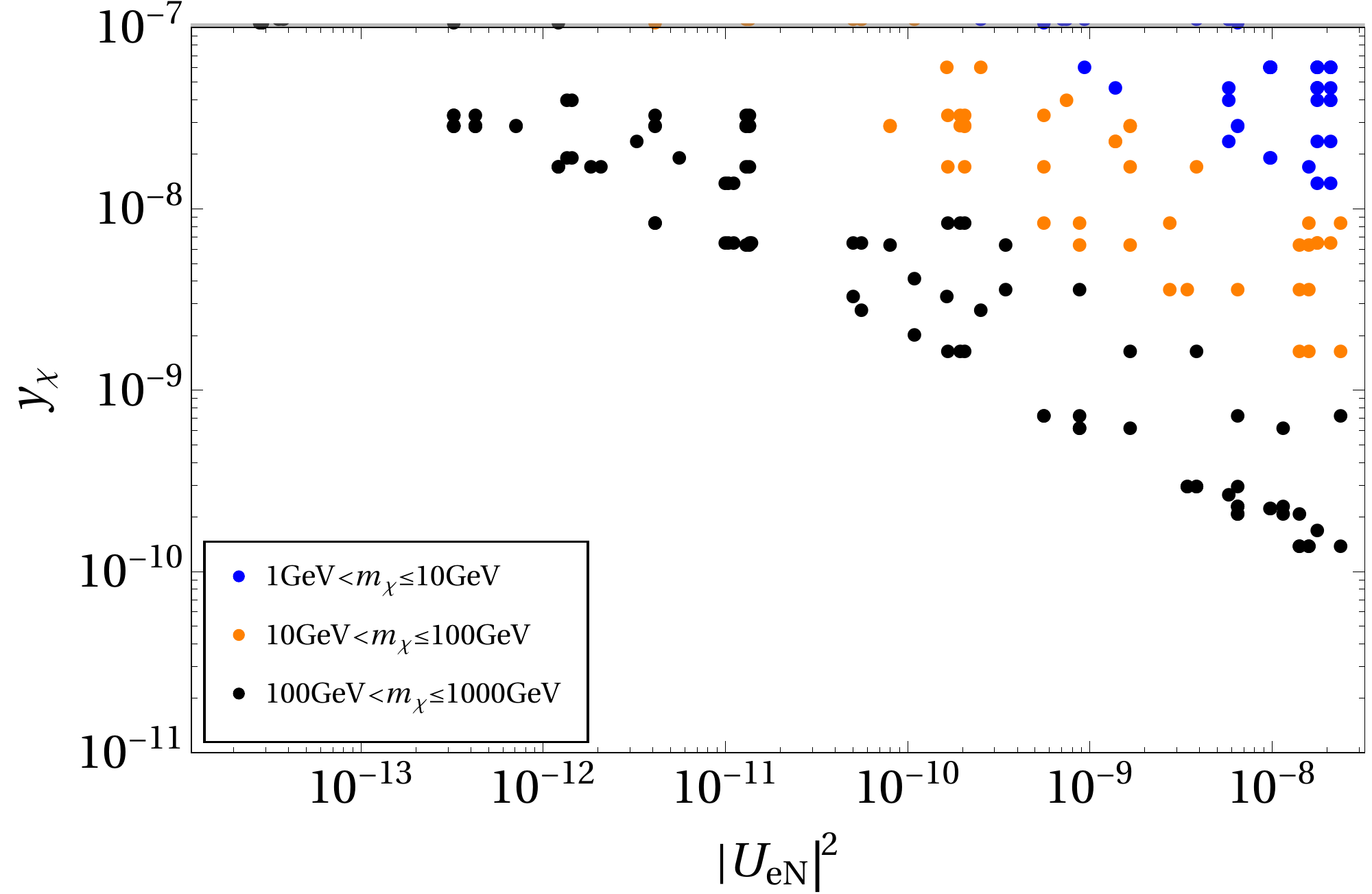}
    \includegraphics[width=0.495\textwidth]{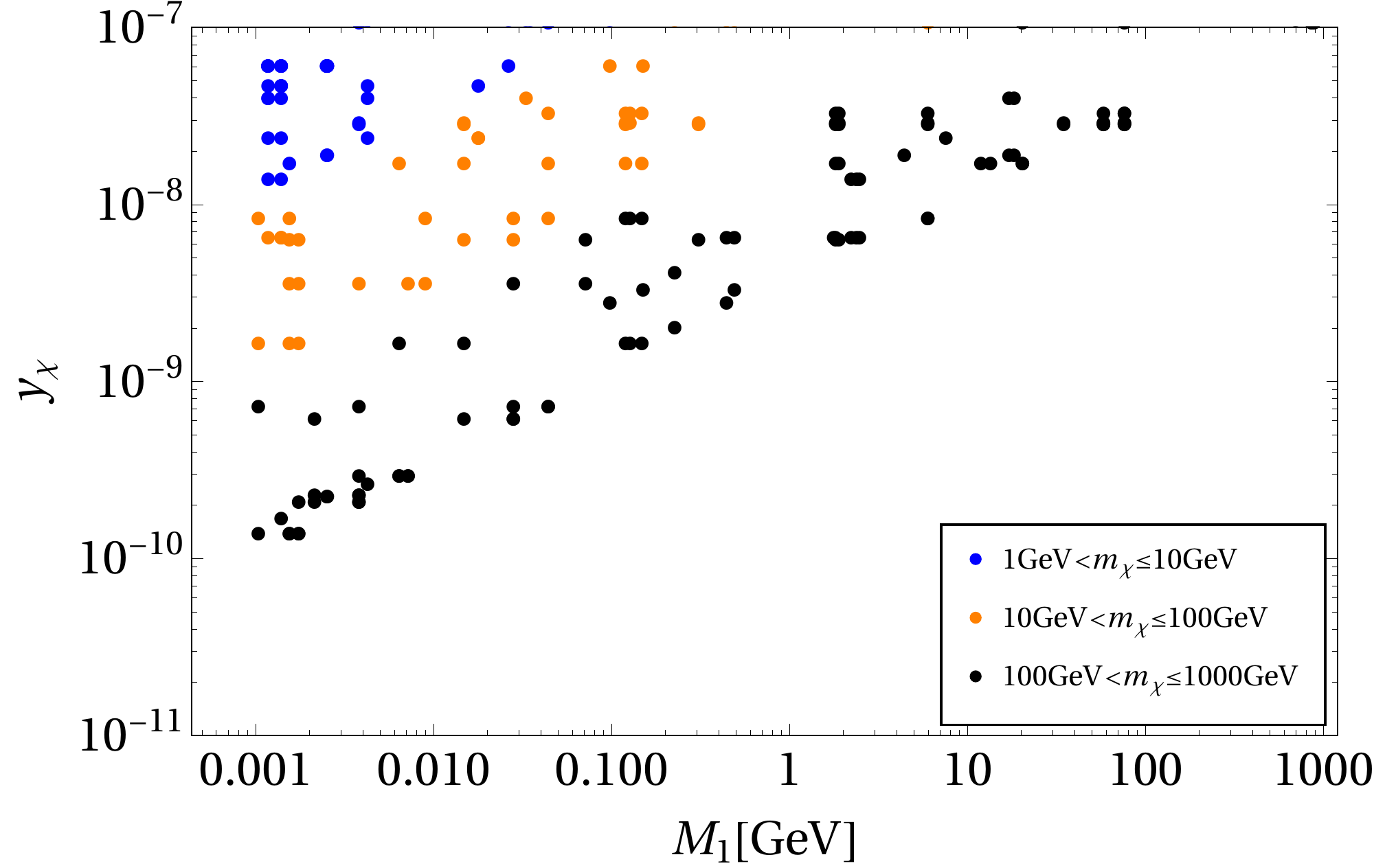}\\[10pt]
    \includegraphics[width=0.495\textwidth]{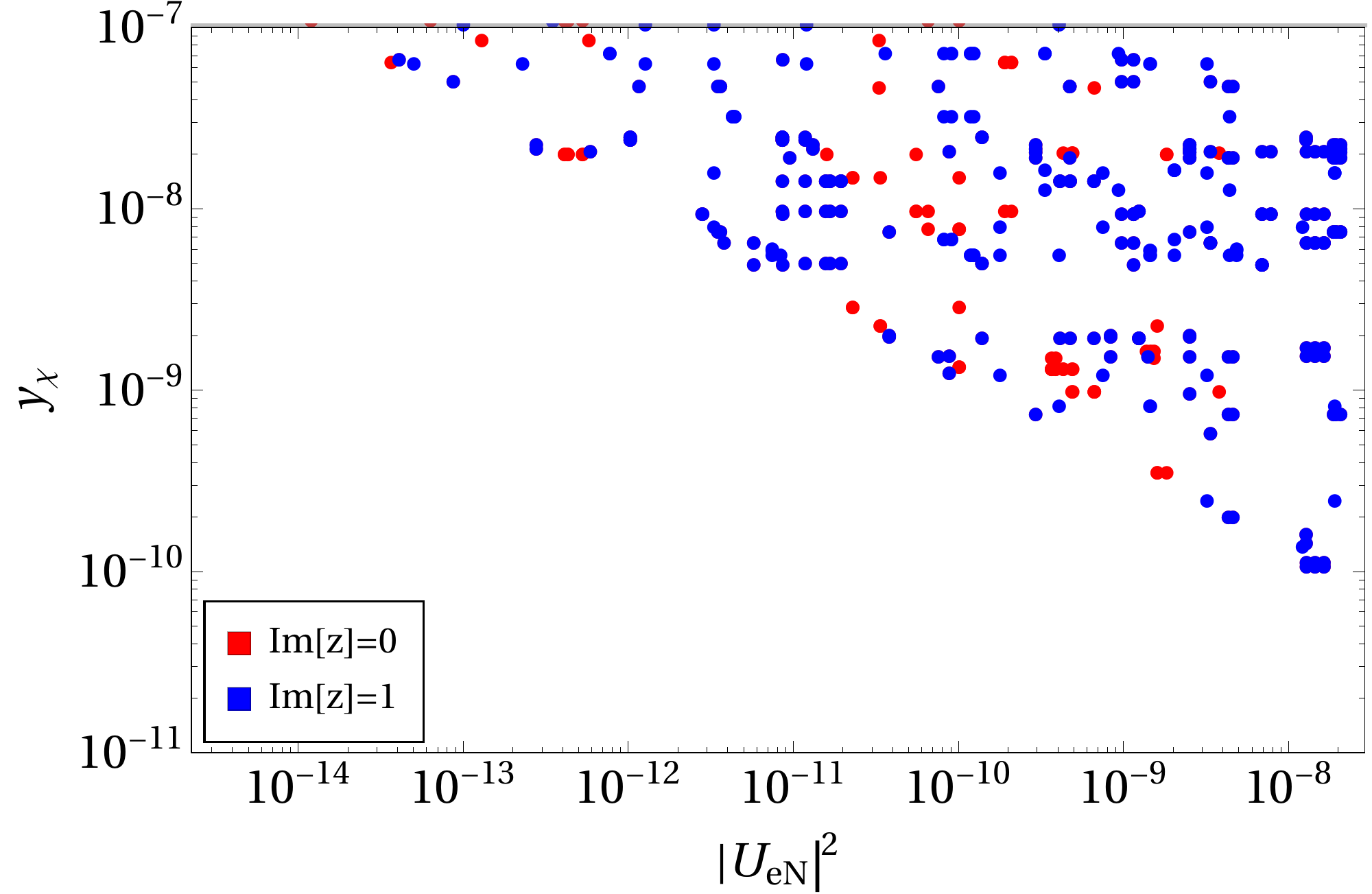}
    \caption{{\it Top Left:} Relic density allowed parameter space in the $(M_1,m_\chi)$ plane where the two colors correspond to different ranges for $y_\chi$ as shown in the plot. {\it Top Right:} Allowed parameter space in the $(m_\chi,y_\chi)$ plane, where different colors denote different RHN mass ranges. {\it Middle Left:} Allowed parameter space in the $(|U_{eN}|^2,y_\chi)$ plane. {\it Middle Right:} Allowed parameter space in the $(M_1,y_\chi)$ plane.  {\it Bottom Center:} Allowed parameter space in the $(|U_{eN}|^2,y_\chi)$ plane, where the two different colors are for $\text{Im}[z]=0$ (red) and $\text{Im}[z]=1$ (blue).}
    \label{fig:relscan2}
\end{figure}

We explore the relic density allowed parameter space for the DM in Fig.~\ref{fig:relscan2}. in the top left panel, the scattered points show the parameter space satisfying the observed relic abundance in the $(M_1,m_\chi)$ plane for $\text{Im}[z]=1$, where the two different colored points indicate different ranges for the $y_\chi$ values. Note that for larger $m_\chi$ a smaller $y_\chi$ is required for satisfying the relic bound as $\Omega_\chi\,h^2\propto y_\chi^2\,m_\chi$. In the top right panel we project the allowed parameter space in the $(m_\chi,y_\chi)$ plane also for $\text{Im}[z]=1$, where different colored points correspond to different $M_1$ mass range. For a fixed DM mass as we increase $y_\chi$, we need to go to larger $M_1$ to satisfy the observed abundance, as that corresponds to smaller $y_N$ [cf.~Eq.~\eqref{eq:umatrix}]. The same is true for a fixed $y_\chi$ as we go to heavier DM mass since in that case over abundance can be avoided by choosing smaller $y_N$, i.e., heavier $M_1$. This feature is more prominent for larger $y_\chi$. As we increase the DM mass, all points tend towards smaller $y_\chi$ to satisfy the relic bound. It is important to note here that for the chosen range of $y_\chi$, the DM is always safe from WDM limit unlike the case of decay. We illustrate the viable DM parameter space in the coupling plane in the remaining panels of Fig.~\ref{fig:relscan2}. The middle left panel shows the $(|U_{eN}|^2,y_\chi)$ plane for different choices of the DM mass and fixing  $\text{Im}[z]=1$. Here we see that small values of $y_\chi$ require heavier DM to satisfy the relic density constraint as $\Omega_\chi\,h^2\propto m_\chi\,(y_\chi\,y_N)^2$. On the other hand, lowering $y_\chi$ moves the parameter space towards right, i.e., larger $|U_{eN}|^2$ to compensate the under abundance accordingly. We see the exact opposite behavior in the middle right panel for the $(M_1,y_\chi)$ parameter space as $U_{\nu N}\propto M_N^{-1/2}$. Finally, in the bottom panel, we show the relic density in the $(|U_{eN}|^2,y_\chi)$ plane for two different values of $\text{Im}[z]$. For small $\text{Im}[z]$, the allowed points shift to smaller values of $|U_{e N}|^2$. As expected, similar behavior also holds in case of $|U_{\mu N}|^2$ and $|U_{\tau N}|^2$, hence we do not show them here.

\begin{figure}[t!]
    \centering
    \includegraphics[width=0.45\textwidth]{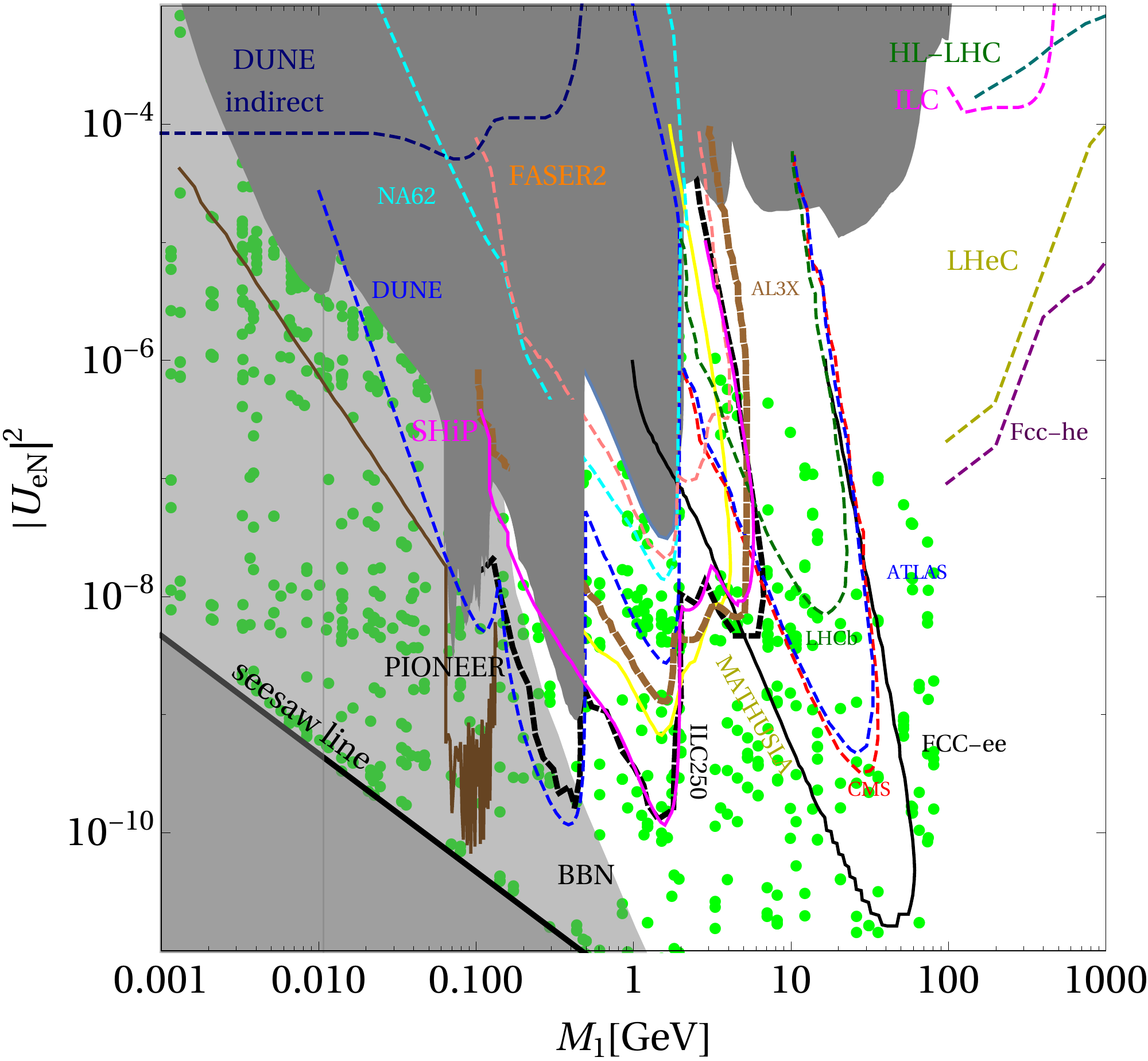}
    \includegraphics[width=0.45\textwidth]{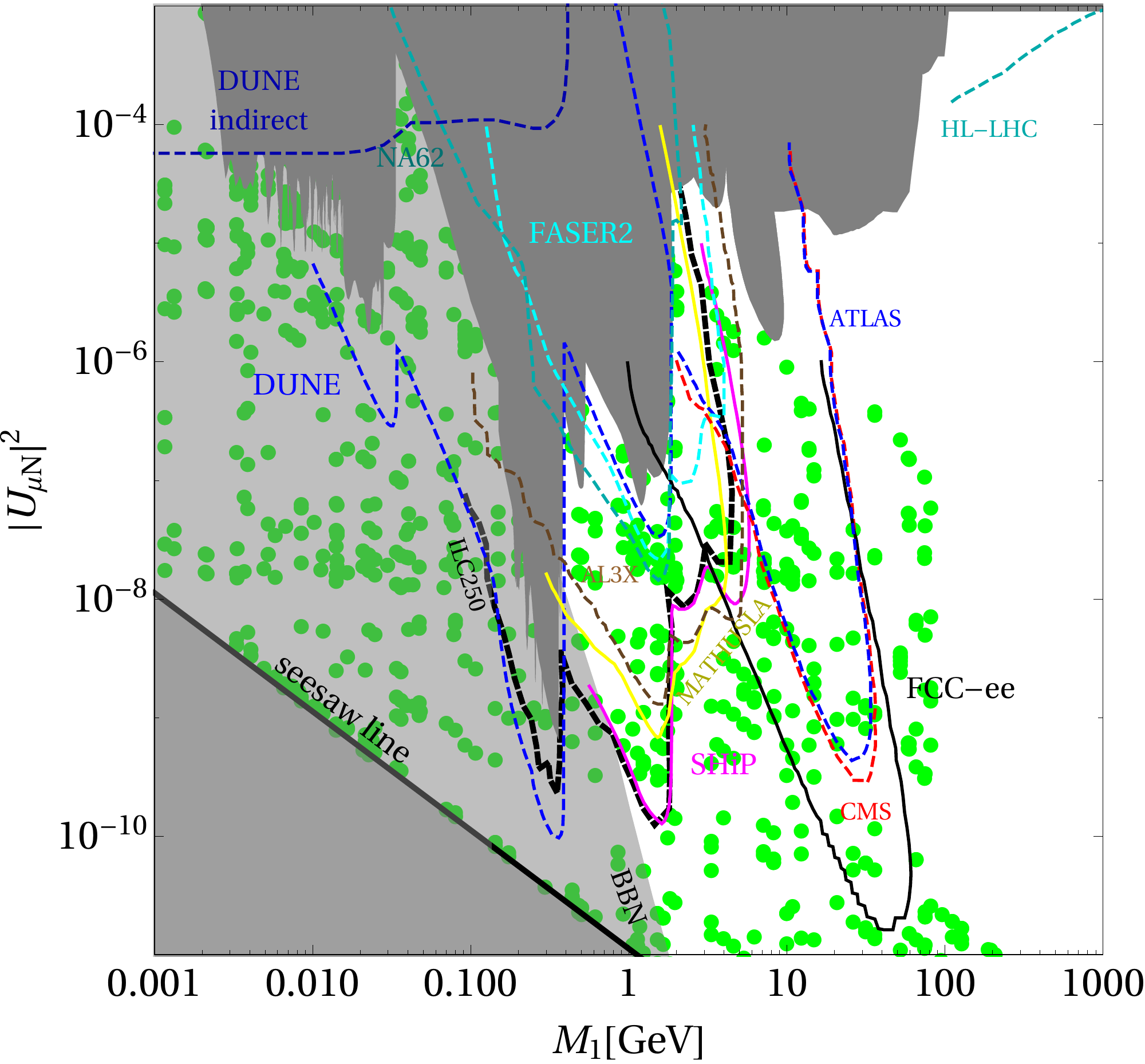}\\[10pt]
    \includegraphics[width=0.45\textwidth]{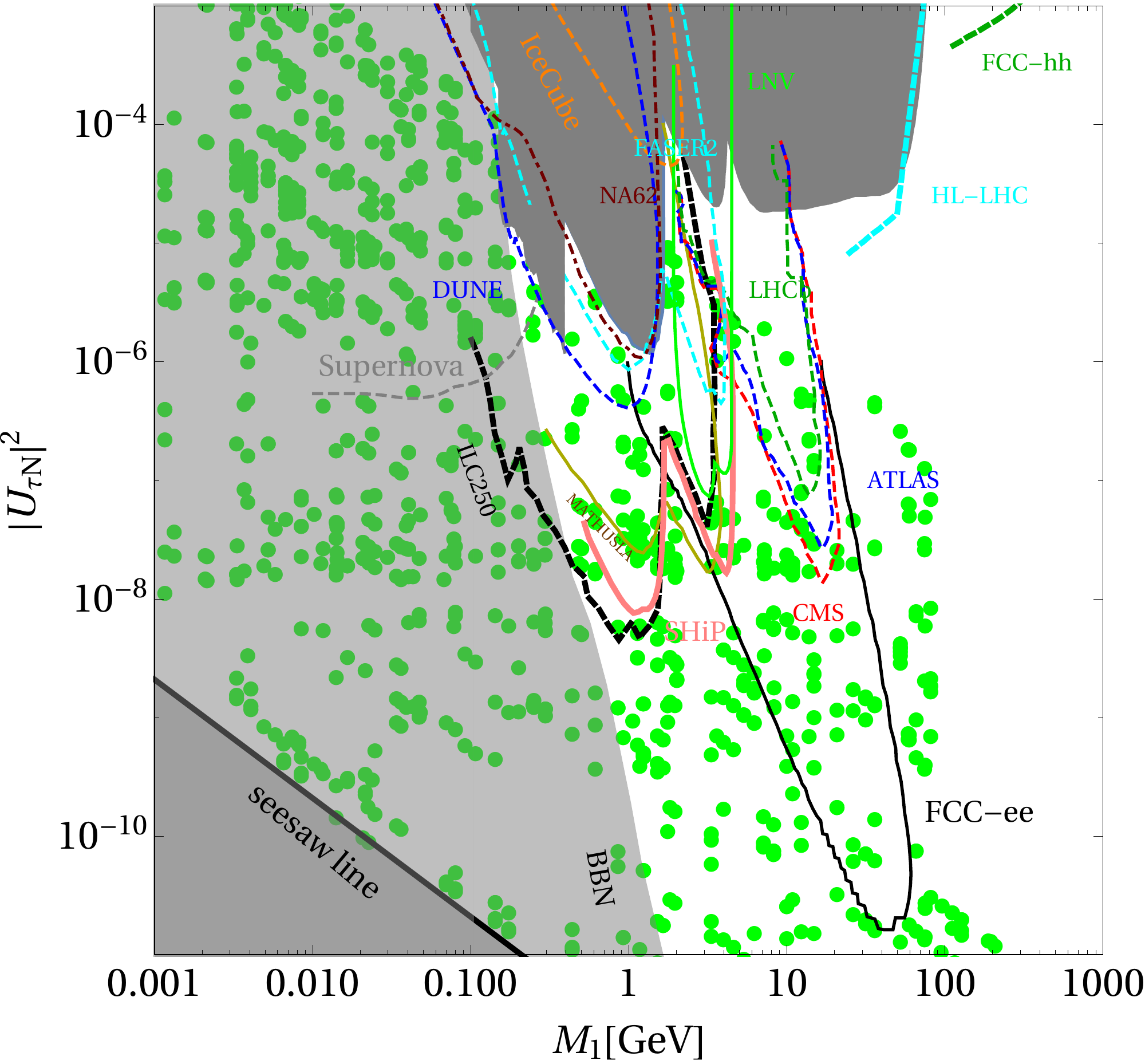}
    \caption{Similar to Fig.~\ref{fig:expt-decay} but for the 2-to-2 scattering case. Here we have considered NH and fixed $y_\chi=10^{-7}$. The dashed black line corresponds to the canonical seesaw case with $\text{Im}[z]=0$.}
    \label{fig:exptlim-22}
\end{figure}
We now cast the allowed DM relic density parameter space in the 2-to-2 scattering case onto the RHN parameter space in Fig.~\ref{fig:exptlim-22}. We project all the relevant limits in $\left|U_{\alpha N}\right|^2-M_1$ as we did in case of decay [cf. Fig.~\ref{fig:expt-decay}]. Here we find that the viable parameter space gradually diminishes towards the right, making a wedge-like shape. This is attributed to the fact that in order to have dominant contribution from scattering, we confine ourselves only in the region of the parameter space where $M_1<m_\chi$, which kinematically forbids the decay channel since we have chosen $m_\varphi=2\,m_\chi$. Another important feature is the absence of the constraint on WDM due to Ly-$\alpha$ here. This is due to the freedom of choosing larger $y_\chi$ in this case, that can still reproduce the correct abundance with a smaller $y_N$, without lowering the DM mass. In case of decay this was not possible as the only coupling controlling the DM abundance was $y_\chi$, hence a larger $y_\chi$ resulted in a lighter DM, making the WDM bound more stringent. 

We should also mention here the possibility that the Yukawa couplings of RHNs can carry new sources of CP violation, and their out-of-equilibrium decays can produce a lepton asymmetry, which can then be transferred to a baryon asymmetry by the electroweak sphalerons -- this is the standard leptogenesis mechanism~\cite{Fukugita:1986hr}. Although the vanilla leptogenesis requires RHN mass to be way above the electroweak scale $(\gtrsim 10^9\,\text{GeV})$~\cite{Davidson:2002qv},  low-scale leptogenesis with RHNs accessible to laboratory experiments is also feasible, either via resonant leptogenesis~\cite{Pilaftsis:2003gt,Dev:2017wwc} or via RHN oscillations~\cite{Akhmedov:1998qx,Drewes:2017zyw} or both~\cite{Klaric:2020phc}. Therefore, in the present scenario it is possible to have a simultaneous explanation of DM relic density and baryon asymmetry since both rely on the interactions of the RHNs. For instance, as has already been discussed in Refs.~\cite{Klaric:2020phc, Hernandez:2022ivz}, it is possible to have successful leptogenesis considering two nearly degenerate RHNs in the GeV range, with $\left|U\right|^2=\sum_\alpha\left|U_\alpha N\right|^2\sim\mathcal{O}(10^{-10}-10^{-8})$, which is also consistent with the DM relic density as shown in Figs.~\ref{fig:expt-decay} and \ref{fig:exptlim-22}. 

\section{The fate of $\varphi$}
\label{sec:phi}
For completeness, let us discuss what happens to the new scalar singlet $\varphi$ in this model. First of all, note that $\varphi$ is non-thermally produced along with the DM $\chi$ unavoidably from $N_1$ decay or 2-to-2 scattering mediated by $N_1$ [cf.~Fig.~\ref{fig:DM-prod}]. On the other hand, $\varphi$ can also be produced from the 2-to-2 scattering of the bath particles due to the presence of the portal interaction $|H|^2\,\varphi^2$ [cf.~Eq.~\eqref{eq:pot}]. This leads to $\varphi$ production via contact interaction before electroweak symmetry breaking, and also via $s$-channel Higgs mediation once the electroweak symmetry is broken and all the SM states become massive (see Appendix~\ref{app:phi-prod}). For the range of masses we are interested in $\varphi$ can be as light as $\mathcal{O}$(MeV) and can be produced via on-shell decay of the Higgs after electroweak symmetry breaking. In that scenario, a large portal coupling $\lhp$ will then contribute to the Higgs invisible branching ratio ${\rm BR}(h\to \varphi\varphi)$, and from the most stringent LHC constraint on $\text{BR}\left(h\to\text{invisible}\right) < 0.145$~\cite{ATLAS:2022yvh} (see also Ref.~\cite{CMS:2022qva} for a slightly weaker bound of 0.18), we find an upper bound of $\lhp\lesssim 6\times 10^{-3}$ for $m_\varphi<m_h/2$. This is weaker than the constraint of $\lhp\lesssim 10^{-7}$ required to keep the $\varphi\varphi\to hh$ process out-of-equilibrium. 

\begin{figure}[t!]
    \centering
    \includegraphics[scale=0.5]{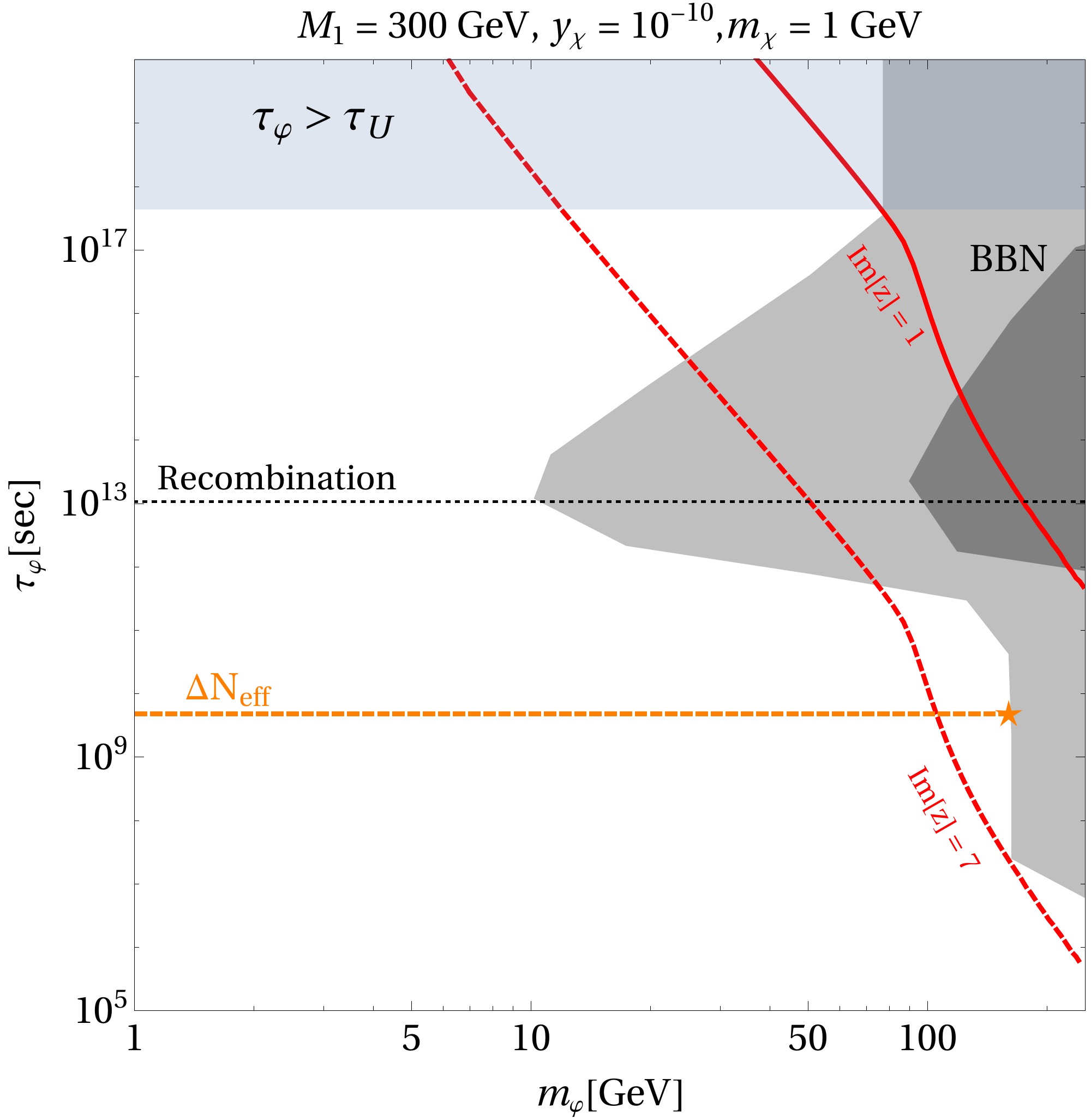}
    \caption{ Lifetime of $\varphi$ considering decays into SM and DM final states via off-shell $N_1$, for $\text{Im}[z]=1$ (solid red) and $\text{Im}[z]=7$ (dashed red). The shaded regions correspond to bounds on $\tau_\varphi$ from CMB+BBN ($\Omega_\varphi/\Omega_\chi=1$ as light gray and $\Omega_\varphi/\Omega_\chi=10^{-2}$ as darker gray), and lifetime of the Universe (pink). The horizontal dashed line shows the $\Delta N_{\rm eff}$ constraint, and the dotted line shows the time of recombination.  All other relevant parameters are fixed at values mentioned in the plot label.}
    \label{fig:3-body}
\end{figure}

After the $\varphi$'s are produced in the early Universe, the next question is about their stability (lifetime).  For $m_\varphi>m_\chi$, $\varphi$ always decays, via the portal coupling $y_\chi$, into $\chi$, which is our primary DM candidate. But $\varphi$ can also be a (decaying) DM candidate if it lives long enough. For $m_\varphi<M_1$, the only possible decay mode of $\varphi$ is to DM and SM final states via off-shell RHN, {\it viz\,.,} 3-body decays $\varphi\to\chi N^*\to \chi\nu h\,, \chi\nu Z\,, \chi \ell^\pm W^\mp$ for $m_\varphi>m_W$, and 4-body decays for $m_\varphi<m_W$ where the SM bosons are produced off-shell and further decay into SM fermions. The dependence of the decay width on the free parameters can be approximately obtained via dimensional analysis as
\begin{align}
& \Gamma_\varphi^\text{3-body}\sim \left(y_\chi\,Y_D\right)^2\,\frac{m_\varphi^5}{M_1^4} \, , \\
& \Gamma_\varphi^\text{4-body}\sim y_\chi^2\,Y_D^2\,g^2\,(y_f^2)\,\frac{m_\varphi^7}{M_1^4\,m_{W,Z}^2\ (m^2_h)}  \, ,  
\end{align}
where $Y_D$ is determined by Eq.~\eqref{eq:CI}. Here $g$ is the SM $SU(2)_L$ gauge coupling and $y_f$ is the Higgs Yukawa coupling as the 4-body decay to light fermions can take place either via the SM weak gauge bosons $W,Z$ or via the Higgs boson $h$. Thus, the decay width (lifetime) increases (decreases) with $m_\varphi$, while it has a inverse dependence on the RHN mediator mass. We numerically calculate the total decay width of $\varphi$ (taking all possible final states into account) and obtain the corresponding lifetime $\tau_\varphi\equiv\Gamma_\varphi^{-1}$ as a function of $m_\varphi$, as plotted in Fig.~\ref{fig:3-body} for two different choices of ${\rm Im}[z]$.

Now, the decay of $\varphi$ into SM final states results in the production of electromagnetic and hadronic showers of particles that can affect the CMB and BBN data. Following Ref.~\cite{Hambye:2021moy}\footnote{Similar studies can also be found, for example, in Refs.~\cite{Poulin:2016nat, Poulin:2016anj, Acharya:2019owx, Acharya:2019uba}.}, we show constraints on $\varphi$ mass and lifetime appearing from energy injection into the CMB photon fluid that impacts the overall CMB power spectra, distortion of the CMB photon spectrum from a pure black body shape (for the same energy injection) and  late-time photodisintegration reactions
that can destroy the predictions from BBN. We show these bounds in Fig.~\ref{fig:3-body} for two different choices of the fractional abundance: $\Omega_{\varphi}/\Omega_{\chi}=1$ (light gray) and $10^{-2}$ (dark gray), and for fixed values of $M_1$, $y_\chi$ and $m_\chi$ as shown in the plot.  The $\Delta N_{\rm eff}$ constraint is also shown for comparison, which is stronger than other CMB and BBN constraints for lighter $\varphi$ masses. 
The horizontal shaded region denotes the $\varphi$ lifetime being larger than the age of the Universe, $\tau_U$, which makes it cosmologically stable. 
As one can see from Fig.~\ref{fig:3-body}, the BBN+CMB constraints completely rule out the $\text{Im}[z]=1$ scenario if we assume the fractional abundance of $\varphi$ to be one, while higher values of $\text{Im}[z]$ still survive. However, for sub-dominant contribution of $\varphi$ to DM abundance, even the $\text{Im}[z]=1$ case is allowed, depending on the $\varphi$ mass. We also checked that changing the RHN or DM mass, and the DM coupling, do not affect our results drastically.

\begin{figure}[t!]
    \centering
    \includegraphics[scale=0.37]{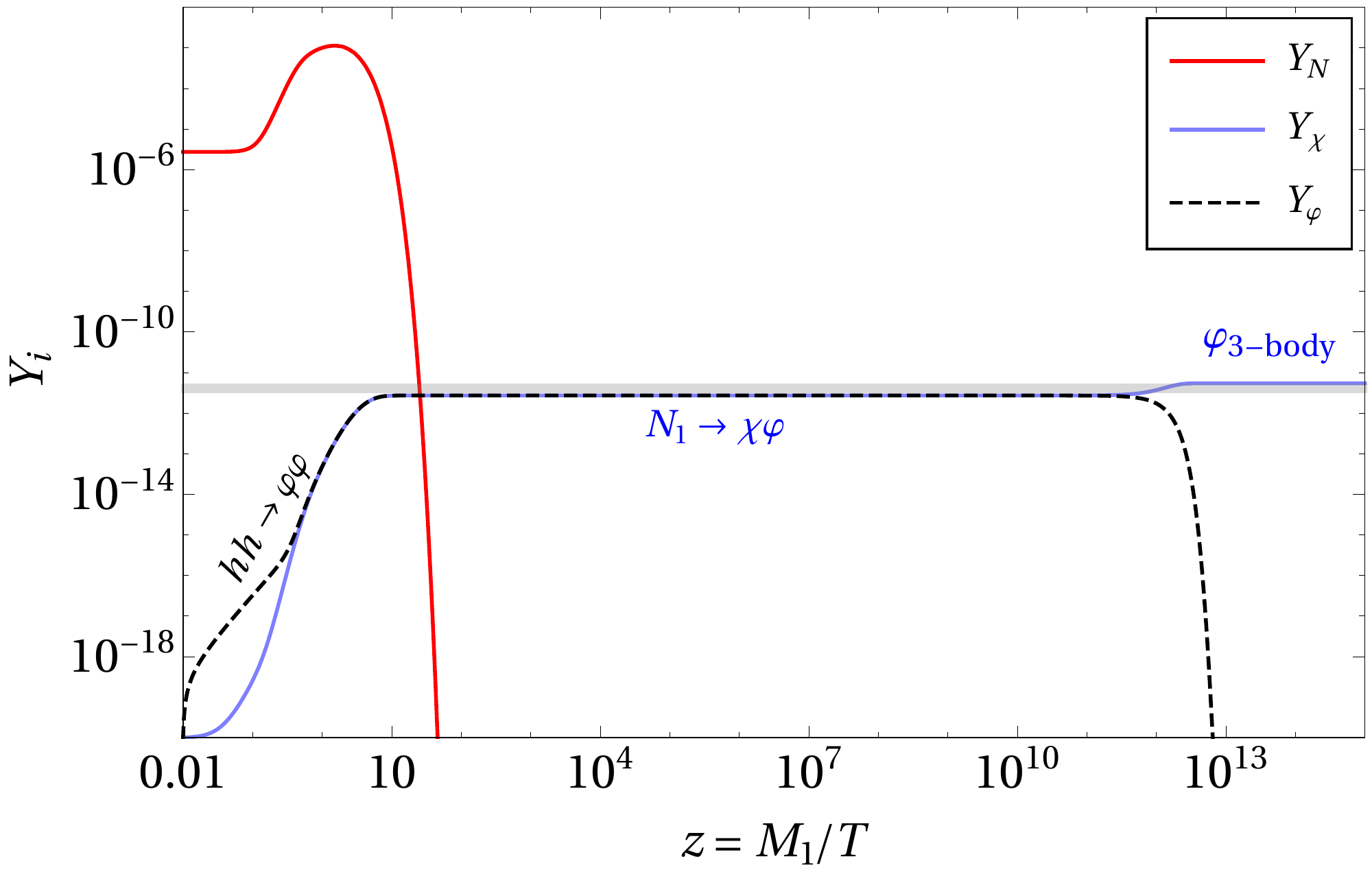}~~\includegraphics[scale=0.37]{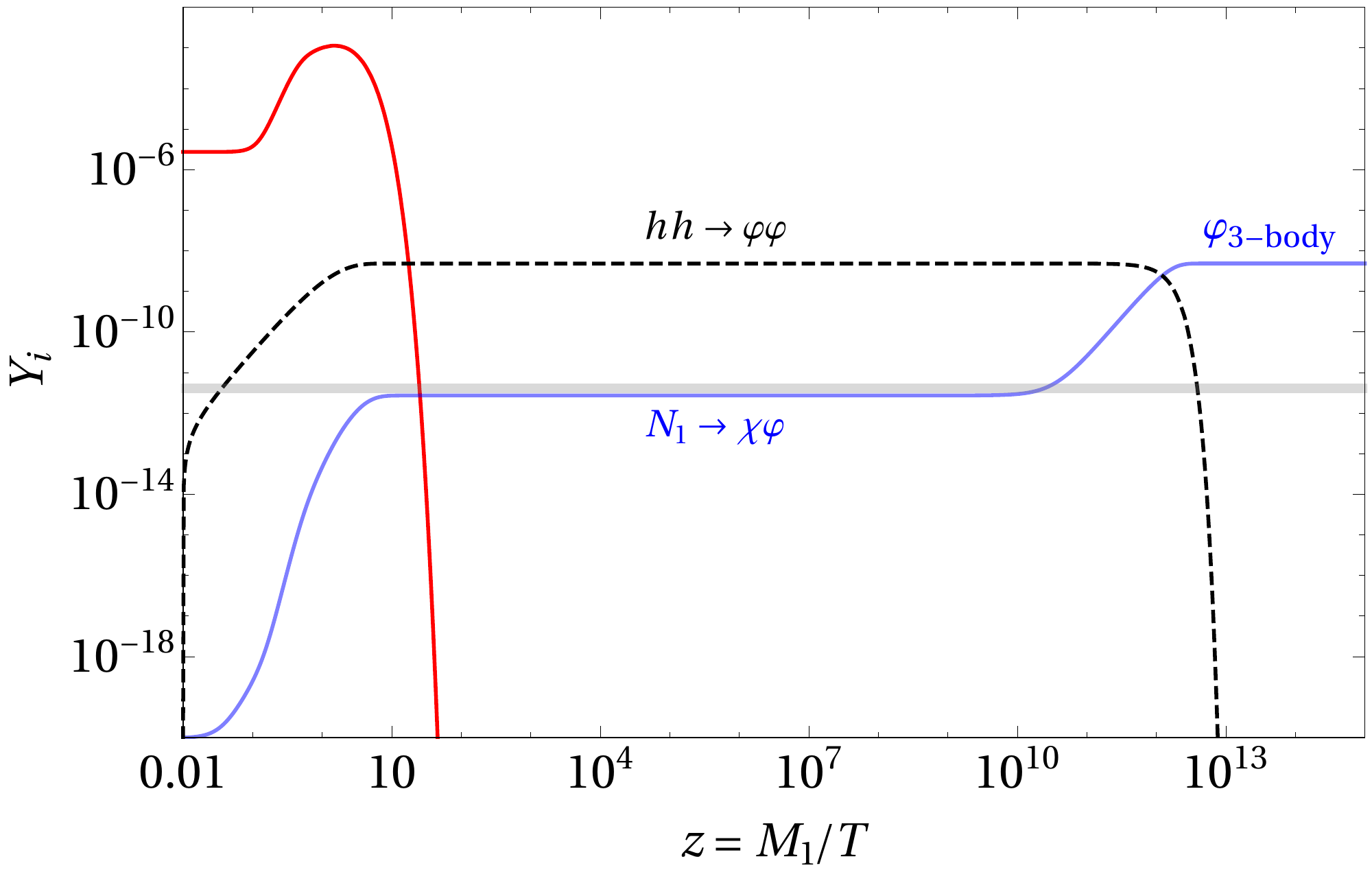}
    \caption{{\it Left:} Evolution of yields $Y_i=n_i/s$, with $i\in N_1\,,\chi\,,\varphi$ with $z=M_1/T$. {\it Right:} Same as left but for $\lambda_{H\varphi}=10^{-9}$. Here we fix $\{M_1\,,m_\chi\,,m_\varphi\}=\{500\,,100\,,300\}$ GeV and $y_\chi=5\times 10^{-12}$, with $\lambda_{H\varphi}=10^{-12}$ (left panel) and $\lambda_{H\varphi}=10^{-9}$ (right panel). The gray shaded horizontal band is where the right DM relic abundance is produced for $\chi$.}
    \label{fig:yld}
\end{figure}

As for the final abundance of $\varphi$ and whether it contributes to the DM abundance, we show in Fig.~\ref{fig:yld} evolution of yields of different components with $z=M_1/T$ in Fig.~\ref{fig:yld}, where we have included contributions to $\varphi$ yield from the SM bath and contributions to $\chi$ yield from late decay of $\varphi$. For the relevant Boltzmann equations, see e.g., Ref.~\cite{Bandyopadhyay:2020qpn}. We fix the parameters in such a way that the right relic abundance is produced for $\chi$. The $N_1$ yield (in red) follows equilibrium distribution, while the yields of $\varphi$ (in black) and $\chi$ (in blue) slowly build up from initial zero abundance. The initial $\varphi$ abundance crucially depends on the Higgs-portal coupling $\lhp$. In the left panel of Fig.~\ref{fig:yld}, we have chosen $\lhp=10^{-12}$ for which the contribution from $hh\to\varphi\varphi$ dominates over the contribution form $N_1\to\chi\,\varphi$ to the initial abundance of $\varphi$, as one can see from the difference in the black dashed and blue curve for $z\lesssim 1$. But soon afterwards, as the temperature drops, the $N_1\to \chi\varphi$ is the only dominant process that contributes equally to both $\chi$ and $\varphi$ abundance.\footnote{The $NN\to \varphi\varphi$ and $NN\to \chi\chi$ processes are suppressed due to an additional power of the Yukawa coupling $y_\chi$ at the amplitude level.} The $\varphi$ abundance eventually goes down once the 3-body decay of $\varphi\to\chi\,\text{SM}\,\text{SM}$ (via off-shell $N_1$) is switched on, where SM includes SM gauge bosons, leptons and Higgs (everything that $N_1$ can decay into via mixing). This happens at $z=M_1/T\sim 10^{12}$ (corresponding to a redshift of $\sim 2200$, twice above the recombination epoch), marked by $``\varphi_\text{3-body}"$. At the same time the $\chi$ abundance also freezes in. Thus for $\lhp\lesssim 10^{-12}$, the contribution to $\chi$ abundance from late decay of $\varphi$ is negligibly small and can affect the $\chi$ yield only at a sub-percentage level. 

For the same set of parameters, except for a larger $\lhp$, as shown in the right panel, leads to larger production of $\varphi$ that results in overabundance of $\chi$ from the late decay of $\varphi$. Correct relic abundance of $\chi$ can still be produced by tuning the other model parameters (such as $M_1\,,\mdm$ or $\ydm$) accordingly. The same is true for the onset of freeze-in, which can always be arranged to happen before recombination, so as to satisfy the cosmological DM constraints. 

\section{Other possible signatures}
\label{sec:collider}

The standard collider signatures of the RHNs~\cite{Atre:2009rg,Deppisch:2015qwa,Das:2016hof,Das:2018usr,Cai:2017mow,Abdullahi:2022jlv}, such as the dilepton (or trilepton) plus missing transverse energy, either prompt or displaced depending the Dirac Yukawa couplings, are also applicable in our scenario. Moreover, the presence of the singlet scalar $\varphi$, in addition to the RHNs, opens up the possibility of distinguishing this model at colliders from the pure seesaw models. For $M_1>m_\varphi+m_\chi$, the branching ratios (BRs) of the RHNs into the SM final states, i.e. $N\to \ell W$, $\nu Z$, $\nu h$, get modified due to the presence of the additional decay mode $N\to \chi\varphi$, depending on the new Yukawa coupling $y_\chi$. Note that the same Yukawa coupling $y_\chi$ governs the DM production in our FIMP scenario; therefore, measuring the BRs accurately can in principle give us a direct collider probe of the RHN coupling to the DM. However, it turns out to be extremely difficult in practice. The reason is that for the decay case, the coupling $y_\chi$ is required to be very small, $\lesssim 10^{-10}$  (cf.~Sec.~\ref{sec:decay}). For comparison, the RHN couplings to the SM fermions, governed by the Dirac Yukawa couplings, are typically of order of $10^{-6}$ for the canonical seesaw case (for a 100 GeV-scale RHN), and can be much larger for larger ${\rm Im}[z]$. Therefore, the RHN decay into the SM final states, either two-body (for $M_N>m_W$) or three-body (for $M_N<m_W$) are always expected to be dominant over the new channel $N\to \chi\varphi$. We have numerically checked that for the parameter space we are interested in here, ${\rm BR}(N\to \chi\varphi)$ can at most of of order of $10^{-5}$ for lighter RHNs, and much smaller for heavier RHNs. Therefore, we do not expect any observable excess in the RHN invisible decay mode (due to $N\to \chi\varphi$) over the standard one ($N\to 3\nu$). 

It is also possible to directly produce $\varphi$ at hadron colliders via gluon fusion, $gg\to h\to\varphi\varphi$, using the trilinear coupling $\lambda_{H\varphi}$. This is very similar to a generic SM-singlet scalar search at the LHC; see e.g. Ref.~\cite{Fuchs:2020cmm}.  However, the production cross-section will be heavily suppressed not only because of the loop-induced Higgs production channel, but also due to the requirement that $\lambda_{H\varphi}\lesssim 10^{-5}$ in order to prevent $\varphi$ from coming into thermal equilibrium with the SM plasma (cf.~Sec.~\ref{sec:phi}). Therefore, the collider prospects of $\varphi$, for instance in the monojet channel, are not so promising in our case. 
Before concluding, we note that the freeze-in scenario considered here can in principle also lead to some DM direct detection as well as indirect detection signatures. As for direct detection, it will be induced by a loop-induced effective DM coupling to the $Z$-boson~\cite{Becker:2018rve, Hambye:2018dpi}. However, for the $y_\chi$ values considered here, the corresponding direct detection cross-sections turn out to be many orders of magnitude below the current constraints~\cite{LUX-ZEPLIN:2022qhg}. As for indirect detection, promising prospects were discussed in Ref.~\cite{Batell:2017rol} for the freeze-out scenario; this however is not the case for our freeze-in scenario where the DM annihilation process is not that efficient, simply because of the huge suppression of the DM number density by the factor of $n_\chi/n_\chi^{\rm eq}$ as compared to the freeze-out case.  On the other hand, the neutrino flux from $\varphi$ decay might be accessible in high-energy neutrino experiments~\cite{Bandyopadhyay:2020qpn}. 

\section{Conclusion}
\label{sec:concl}
In this work we have looked into the possibility of probing freeze-in DM coupling via the heavy neutrino portal. We have minimally extended the type-I seesaw scenario with the addition of a gauge-singlet fermion $\chi$ and a real singlet scalar $\varphi$. Both $\varphi$ and $\chi$ are considered to be odd under some stabilizing $Z_2$ symmetry and the fermion $\chi$ is considered to be the viable DM candidate given $m_\chi<m_\varphi$. The DM only talks to the SM sector via its coupling to the RHNs of the form $y_\chi N\chi \varphi$. Depending on whether $M_N$ is lighter or heavier than the sum of $m_\chi$ and $m_\varphi$, the DM can be produced non-thermally either from the decay of the RHNs, considered to be part of the thermal bath, or via 2-to-2 scattering of the SM particles mediated by the RHNs. This is referred to here as the heavy neutrino-portal freeze-in. Using the Casas-Ibarra parametrization to satisfy the neutrino oscillation data with two RHNs, we are left with four free parameters of the model: the DM mass $m_\chi$, DM Yukawa coupling with the RHN $y_\chi$, mass of the new singlet scalar $m_\varphi$ and the lightest RHN mass $M_1$.  

We find that the requirement of freeze-in production of DM (together with the Planck-observed relic abundance) necessarily requires the DM Yukawa coupling $y_\chi\lesssim 10^{-10}$ (Fig.~\ref{fig:rate}) in case the DM is produced from the on-shell decay of the RHN, while for scattering this bound can be significantly relaxed to $y_\chi\lesssim 10^{-7}$ (Fig.~\ref{fig:22rate}) because of the involvement of RHN Yukawa couplings with the SM. Assuming sizable active-sterile neutrino mixing with RHN mass lying in the MeV-TeV range, our RHN portal  scenario can fall within the reach of several current and future facilities, including collider, beam dump and forward physics experiments, that typically look for feebly-coupled heavy neutrinos, as shown in Fig.~\ref{fig:expt-decay} and Fig.~\ref{fig:exptlim-22}. This, in turn, provides a complementary window to probe the freeze-in DM parameter space.

We finally comment that within this framework it is also possible to accommodate low-scale leptogenesis, leading to a common origin of freeze-in DM and baryon asymmetry that can actually be tested. 

\section*{Acknowledgments}
BB would like to acknowledge useful discussions with Arghyajit Datta and Rishav Roshan, and particularly grateful to Rishav Roshan for helping out with the numerical simulations. AG thanks Marco Drewes for discussion regarding the constraints of RHN searches. BB received funding from Patrimonio Autónomo - Fondo Nacional de Financiamiento para la Ciencia, la Tecnología y la Innovación Francisco José de Caldas (MinCiencias - Colombia) grant 80740-465-2020. This project has received funding /support from the European Union's Horizon 2020 research and innovation programme under the Marie Sklodowska-Curie grant agreement No 860881-HIDDeN. The work of BD was supported in part by the U.S. Department of Energy under Grant No.~DE-SC0017987, by a Fermilab Intensity Frontier Fellowship, and by a URA VSP Fellowship. BD acknowledges the Fermilab theory group for local hospitality, where part of this work was performed. 

\appendix
\section{Reaction densities}
\label{sec:app-reacden}

The reaction density corresponding to $1\to2$ decay process is given by
\begin{align}
\gamma_\text{decay} &=\int\,\prod_{i=1}^3\,(2\,\pi)^4\,\delta^{(4)}\,\left(p_a-p_1-p_2\right)\,f_a^\text{eq}\,\left|\mathcal{M}\right|^2_{a\to1,2}\nonumber\\&=\frac{g_a}{2\,\pi^2}\,m_a^2\,\Gamma_{a\to1,2}\,T\,K_1\left(\frac{m_a}{T}\right)\,. 
\end{align}

For 2-to-2 processes the reaction density reads
\begin{equation}\begin{aligned}
\gamma_\text{ann}&=\int\prod_{i=1}^4 d\Pi_i \left(2\pi\right)^4 \delta^{(4)}\biggl(p_a+p_b-p_1-p_2\biggr)f_a{^\text{eq}}f_b{^\text{eq}}\left|\mathcal{M}_{a,b\to1,2}\right|^2\\&=\frac{T}{32\pi^4}\,g_a g_b\,\int_{s_\text{min}}^\infty ds\,\frac{\biggl[\bigl(s-m_a^2-m_b^2\bigr)^2-4m_a^2 m_b^2\biggr]}{\sqrt{s}}\,\sigma\left(s\right)_{a,b\to1,2}\,K_1\left(\frac{\sqrt{s}}{T}\right)\label{eq:gam-ann}\,, \end{aligned}    
\end{equation}
with $a,b(1,2)$ as the incoming (outgoing) states and $g_{a,b}$ are corresponding degrees of freedom. Here $f_i{^\text{eq}}\approx\exp^{-E_i/T}$ is the Maxwell-Boltzmann distribution. The Lorentz invariant 2-body phase space is denoted by: $d\Pi_i=\frac{d^3p_i}{\left(2\pi\right)^3 2E_i}$. The amplitude squared (summed over final and averaged over initial states) is denoted by $\left|\mathcal{M}_{a,b\to1,2}\right|^2$ for a particular 2-to-2 scattering process. The lower limit of the integration over $s$ is $s_\text{min}=\text{max}\left[\left(m_a+m_b\right)^2,\left(m_1+m_2\right)^2\right]$.

\section{RHN decay widths}
\label{sec:app-RHN-decay}
\subsection*{2-body decays:}
\begin{align}\label{eq:rhn-2body}
&\Gamma_{N_i\to\ell_j W^\pm}=\frac{g^2\,M_N^3}{64\,\pi\,M_W^2}\left|\left(U_{\nu N}\right)_{ij}\right|^2\,\left(1-3\,x^4+2\,x^6\right)\,;\,x=M_W/M_N \, ,
\nonumber\\&
\Gamma_{N_i\to\nu_j Z}=\frac{g^2\,M_N^3}{64\,\pi\,M_W^2}\left|\left(\pmns^\dagger\,U_{\nu N}\right)_{ij}\right|^2\,\left(1-3\,x^4+2\,x^6\right)\,;\,x=M_Z/M_N \, ,
\nonumber\\&
\Gamma_{N_i\to\nu_j h}=\frac{g^2\,M_N^3}{64\,\pi\,M_W^2}\,\left|\left(\pmns^\dagger\,U_{\nu N}\right)_{ij}\right|^2\,\left(1-x^2\right)^2\,;\,x=m_h/M_N\,,
\nonumber\\&
\Gamma_{N_i\to\varphi\chi}=\frac{\ydm^2\,M_N}{16\,\pi}\left|U_{NN}\right|^2\,\left[(1+x)^2-y^2\right]\,\sqrt{(1-x^2)^2-2\,y^2\,(1+x^2)+y^4}\,;\,x=\mdm/M_N\,,y=m_\varphi/M_N\,.
\end{align}

\subsection*{3-body decays (leptonic final states):}

\begin{align}\label{eq:rhn-3bodya}
\Gamma_{\ell_1^-\ell_2^+\nu_{\ell_2}} = & \frac{G_\text{F}^2}{192\,\pi^3}\,M_N^5\,\left|U_{\nu N}\right|^2\,\mathcal{I}(x_{\ell_1},x_{\nu_{\ell_2}},x_{\ell_2})\,;
\nonumber\\
\Gamma_{\ell_2^+\ell_2^-\nu_{\ell_1}} = & \frac{G_\text{F}^2}{96\,\pi^3}\,M_N^5\,\left|\mathcal{X}\right|^2\,\Biggl[\Bigl(g_L\,g_R+\delta_{\ell_1\ell_2}\,g_R\Bigr)\,\mathcal{J}(x_{\nu_{\ell_1}},x_{\ell_2},x_{\ell_2})\nonumber\\& \qquad +\Bigl(g_L^2+g_R^2+\delta_{\ell_1\ell_2}(1+2\,g_L)\Bigr)\,\mathcal{I}(x_{\nu_{\ell_1}},x_{\ell_2},x_{\ell_2})\Biggr]\,;
\nonumber\\
\Gamma_{\nu_{\ell_1}\nu_{\ell_1}\nu_{\ell_2}} =&  \frac{G_\text{F}^2}{96\,\pi^3}\,\left|U^\dagger_\text{PMNS}\,U_{\nu N}\right|^2\,M_N^5\,,
\end{align}
where $\mathcal{X}\equiv \pmns^\dagger\,U_{\nu N} (U_{\nu N})$ for neutral (charged) current ,and
\begin{align}
\mathcal{I}(x,y,z) = & 12\,\int_{(x+y)^2}^{(1-z)^2}\,\frac{ds}{s}\,(s-x^2-y^2)\,(1+z^2-s)\,\lambda(s,x^2,y^2)^{1/2}\,\lambda(1,s,z^2)^{1/2}\,,  \\ 
\mathcal{J}(x,y,z)= & 24\,yz\,\int_{(y+z)^2}^{(1-x)^2}\,\frac{ds}{s}\,(1+x^2-s)\,\lambda(s,y^2,z^2)^{1/2}\,\lambda(1,s,x^2)^{1/2}\,,    
\end{align}
with
\begin{equation}
\lambda(x,y,x)=x^2+y^2+z^2-2xy-2yz-2zx\,,x_i=m_i/M_N\,,
\end{equation}
and $g_L=-\frac{1}{2}+\sin^2\theta_w, g_R=\sin^2\theta_w$, $\theta_w$ being the weak mixing angle. 

\subsection*{3-body decays (semi-leptonic final states):}

\begin{align}\label{eq:rhn-3bodyb}
& \Gamma_{\ell^-P_S^+}=\frac{G_\text{F}^2}{16\,\pi}\,f_P^2\,\left|U_\text{CKM}\right|^2\,\left|U_{\nu N}\right|^2\,M_N^3\,\mathcal{A}(x_\ell\,,x_P)\,;
\nonumber\\&
\Gamma_{\nu_{\ell}P_S^0}=\frac{G_\text{F}^2}{64\,\pi}\,f_P^2\,\,\left|\pmns^\dagger\,U_{\nu N}\right|^2\,M_N^3\,(1-x_P)^2\,;
\nonumber\\&
\Gamma_{\ell^- V^+}=\frac{G_\text{F}^2}{16\,\pi}\,f_V^2\,\left|U_\text{CKM}\right|^2\,\left|U_{\nu N}\right|^2\,M_N^3\,\mathcal{B}(x_\ell\,,x_V)\,;
\nonumber\\&
\Gamma_{\nu_{\ell}V^0}=\frac{G_\text{F}^2}{2\,\pi}\,\kappa_V^2\,f_V^2\,\left|\pmns^\dagger\,U_{\nu N}\right|^2\,M_N^3\,\mathcal{C}(x_{\nu_\ell}\,,x_V)\,;
\end{align}
where $P_S^{0(\pm)}$ are the neutral (charged mesons) and
\begin{align}
&\mathcal{A}(a,b)=\Bigl[(1+a-b)\,(1+a)-4\,a\Bigr]\,\lambda(1,a,b)^{1/2}\,;
\noindent\\&
\mathcal{B}(a,b)=\Bigl[(1+a-b)\,(1+a+2b)-4\,a\Bigr]\,\lambda(1,a,b)^{1/2}\,;
\nonumber\\&
\mathcal{C}(a,b)=(1+2\,b)\,(1-b)\,\lambda(1,a,b)^{1/2}\,,
\end{align}
with $x_i=m_i^2/M_N^2$, $f_i$ are the meson decay constants and $\kappa_V$ being the vector coupling associated with the meson as in Ref.~\cite{Atre:2009rg}. 
\section{2-to-2 cross-sections for freeze-in}
\label{sec:app-22-dm}
The cross-sections are expressed in the limit of zero mass of the SM leptons. 
\subsection*{s-channel:}
\begin{align}
\sigma(s)_{h\nu\to\varphi\chi}=& \frac{y_\chi^2\,\alpha}{256\,s^2}\,\left|\pmns^\dagger\,U_{\nu N}\right|^2\,\Biggl[\frac{M_1^2\,(s-m_h^2)}{M_W^2\,(1-M_W^2/M_Z^2)}\Biggr]\nonumber\\
&\Biggl[\frac{(s+M_1^2) \left(s-m_\varphi^2+m_\chi^2\right)+4\,s\,M_1\,m_\chi}{(s-M_1^2)^2+\Gamma_1^2\,M_1^2}\Biggr]\,\Biggl[\frac{m_\varphi^4-2 m_\varphi^2 \left(s+m_\chi^2\right)+\left(s-m_\chi^2\right)^2}{(s-M_Z^2)^2}\Biggr]^{1/2}\,,
\nonumber\\
\sigma(s)_{Z\nu\to\varphi\chi}=& \frac{y_\chi^2\,\alpha}{768\,s^2\,M_W^2}\,\left|\pmns^\dagger\,U_{\nu N}\right|^2\,\Biggl[\frac{M_Z^2\,(s-M_Z^2)\,(s+2\,M_Z^2)}{(M_Z^2-M_W^2)}\Biggr]
\nonumber\\&
\Biggl[\frac{(s+M_1^2) \left(s-m_\varphi^2+m_\chi^2\right)+4\,s\,M_1\,m_\chi}{(s-M_1^2)^2+\Gamma_1^2\,M_1^2}\Biggr]\,\Biggl[\frac{m_\varphi^4-2 m_\varphi^2 \left(s+m_\chi^2\right)+\left(s-m_\chi^2\right)^2}{(s-M_Z^2)^2}\Biggr]^{1/2}
\nonumber\\
\sigma(s)_{W\ell\to\varphi\chi}= & \frac{y_\chi^2\,\alpha}{384\,s^2\,M_W^2}\,\left|U_{\nu N}\right|^2\,\Biggl[\frac{s^2-2\,M_W^4+s\,M_W^2}{M_Z^2-M_W^2}\Biggr]\,
\nonumber\\&
\Biggl[\frac{(s+M_1^2) \left(s-m_\varphi^2+m_\chi^2\right)+4\,s\,M_1\,m_\chi}{(s-M_1^2)^2+\Gamma_1^2\,M_1^2}\Biggr]\,\Biggl[\frac{m_\varphi^4-2 m_\varphi^2 \left(s+m_\chi^2\right)+\left(s-m_\chi^2\right)^2}{(s-M_W^2)^2}\Biggr]^{1/2}\,.
\end{align}
\subsection*{t-channel:}
\begin{align}
& \sigma(s)_{N_1N_1\to\chi\chi} \simeq \frac{y_\chi^4}{128\,\pi\,s\,(M_1-\mdm)^2\,(s-4 M_1^2)\,\left(s-2\left(M_1^2+\mdm^2\right)\right)}
\nonumber\\&
\left[\left(5M_1^2-2M_1\,\mdm+5 \mdm^2\right) \sqrt{(s-4 \mdm^2)\,(s-4M_1^2)}\,\left(2 \left(s-M_1^2+\mdm^2\right)\right)\right]+\mathcal{O}[m_\varphi^2]\,.
\end{align}

\section{Production cross-sections for $\varphi$}
\label{app:phi-prod}
\begin{align}
& \sigma(s)_{HH\to\varphi\,\varphi}=\frac{\lhp^2}{32\,\pi\,s}\,\sqrt{1-\frac{4\,m_\varphi^2}{s}}
\nonumber\\&
\sigma(s)_{hh\to\varphi\varphi} =  \frac{\lhp^2\,\sqrt{1-\frac{4 m_\varphi^2}{s}}\,\left(s-m_h^2+6\lambda^\text{SM}\,v_h^2\right)^2}{8 \pi  s \sqrt{1-\frac{4 m_h^2}{s}} \left(\left(s-m_h^2\right)^2+\Gamma_h^2\,m_h^2\right)}
\nonumber\\&
\sigma(s)_{ff\to\varphi\varphi} =\frac{\lhp^2 m_f^2 \sqrt{\left(1-\frac{4 m_f^2}{s}\right)\,\left(1-\frac{4m_\varphi^2}{s}\right)}} {48\,\pi\left(\left(s-m_h^2\right)^2+\Gamma_h^2 m_h^2\right)}
\nonumber\\&
\sigma(s)_{VV\to\varphi\varphi} = \frac{\pi \,\alpha^2\,\lhp^2\, v_h^4\,\sqrt{1-\frac{4 m_\varphi^2}{s}}\,\left(s^2-4\,m_V^2\,s+12 m_V^4\right)}{72\,m_W^4\,s\,\left(1-\frac{m_W^2}{m_Z^2}\right)^2 \sqrt{1-\frac{4 m_V^2}{s}} \left(\left(s-m_h^2\right)^2+\Gamma_h^2 m_h^2\right)}
\nonumber\\&
\sigma(s)_{N_1N_1\to\varphi\varphi} \simeq \frac{y_\chi^4}{256 \pi s (M_1-m_\varphi) (M_1+m_\varphi) \left(4 M_1^2-s\right) \left(s-2 \left(M_1^2+m_\varphi^2\right)\right)}
\nonumber\\&
\times \left[2\sqrt{(s-4 M_1^2)\,(s-4 m_\varphi^2)}\left(3 M_1^2+4 M_1 \mdm-3 m_\varphi^2\right) \left(s-2 \left(M_1^2+m_\varphi^2\right)\right)\right]\, , 
\end{align}
where $N_c=1~(3)$ is the color factor for lepton (quark) initial state, $V\in W^\pm,\,Z$ for the weak gauge bosons, $\lambda^\text{SM}\simeq 0.3$ is the SM Higgs quartic coupling, and $\alpha\simeq 1/128$ is the fine structure constant evaluated at the electroweak scale.

\bibliographystyle{utcaps_mod}
\bibliography{Bibliography}


\end{document}